\documentclass[longauth]{aa}
\usepackage[varg]{txfonts} 
\usepackage{amssymb,amsmath}
\usepackage{enumerate}
\usepackage{graphicx}
\usepackage{subfigure}
\usepackage{color}
\usepackage{float}
\usepackage[T1]{fontenc}
\usepackage[english]{babel}
\usepackage{natbib}
\usepackage{booktabs}

\begin{document}
\title{Long-term multi-wavelength variability and correlation study of Markarian~421 from 2007 to 2009}

%
\author{
M.~L.~Ahnen\inst{1} \and
S.~Ansoldi\inst{2} \and
L.~A.~Antonelli\inst{3} \and
P.~Antoranz\inst{4} \and
A.~Babic\inst{5} \and
B.~Banerjee\inst{6} \and
P.~Bangale\inst{7} \and
U.~Barres de Almeida\inst{7,}\inst{24} \and
J.~A.~Barrio\inst{8} \and
J.~Becerra Gonz\'alez\inst{9,}\inst{25}
% ...
}
\institute { ETH Zurich, CH-8093 Zurich, Switzerland
\and Universit\`a di Udine, and INFN Trieste, I-33100 Udine, Italy
\and INAF National Institute for Astrophysics, I-00136 Rome, Italy
\and Universit\`a  di Siena, and INFN Pisa, I-53100 Siena, Italy
\and Croatian MAGIC Consortium, Rudjer Boskovic Institute, University of Rijeka, University of Split and University of Zagreb, Croatia
\and Saha Institute of Nuclear Physics, 1/AF Bidhannagar, Salt Lake, Sector-1, Kolkata 700064, India
\and Max-Planck-Institut f\"ur Physik, D-80805 M\"unchen, Germany
\and Universidad Complutense, E-28040 Madrid, Spain
\and Inst. de Astrof\'isica de Canarias, E-38200 La Laguna, Tenerife, Spain; Universidad de La Laguna, Dpto. Astrof\'isica, E-38206 La Laguna, Tenerife, Spain
\and University of \L\'od\'z, PL-90236 Lodz, Poland
\and Deutsches Elektronen-Synchrotron (DESY), D-15738 Zeuthen, Germany
\and Institut de Fisica d'Altes Energies (IFAE), The Barcelona Institute of Science and Technology, Campus UAB, 08193 Bellaterra (Barcelona), Spain
\and Universit\"at W\"urzburg, D-97074 W\"urzburg, Germany
\and Universit\`a di Padova and INFN, I-35131 Padova, Italy
\and Institute for Space Sciences (CSIC/IEEC), E-08193 Barcelona, Spain
\and Technische Universit\"at Dortmund, D-44221 Dortmund, Germany
\and Unitat de F\'isica de les Radiacions, Departament de F\'isica, and CERES-IEEC, Universitat Aut\`onoma de Barcelona, E-08193 Bellaterra, Spain
\and Universitat de Barcelona, ICC, IEEC-UB, E-08028 Barcelona, Spain
\and Japanese MAGIC Consortium, ICRR, The University of Tokyo, Department of Physics and Hakubi Center, Kyoto University, Tokai University, The University of Tokushima, KEK, Japan
\and Finnish MAGIC Consortium, Tuorla Observatory, University of Turku and Astronomy Division, University of Oulu, Finland
\and Inst. for Nucl. Research and Nucl. Energy, BG-1784 Sofia, Bulgaria
\and Universit\`a di Pisa, and INFN Pisa, I-56126 Pisa, Italy
\and ICREA and Institute for Space Sciences (CSIC/IEEC), E-08193 Barcelona, Spain
\and now at Centro Brasileiro de Pesquisas F\'isicas (CBPF/MCTI), R. Dr. Xavier Sigaud, 150 - Urca, Rio de Janeiro - RJ, 22290-180, Brazil
\and now at NASA Goddard Space Flight Center, Greenbelt, MD 20771, USA and Department of Physics and Department of Astronomy, University of Maryland, College Park, MD 20742, USA
% ...
}

\abstract{}
{We study the multi-band variability and correlations of the TeV blazar Mrk~421 on year time scales, which can bring additional insight on the processes responsible for its broadband emission.}
{We observed Mrk~421 in the very high energy (VHE) $\gamma$-ray range with the Cherenkov telescope MAGIC-I from March 2007 to June 2009 for a total of 96~hours of effective time after quality cuts. The VHE flux variability is quantified with several methods, including the Bayesian Block algorithm, which is applied to data from Cherenkov telescopes for the first time. The 2.3 year long MAGIC light curve is complemented with data from the \textit{Swift}/BAT and \textit{RXTE}/ASM satellites and the KVA, GASP-WEBT, OVRO, and Mets\"ahovi telescopes from February 2007 to July 2009, allowing for an excellent characterisation of the multi-band variability and correlations over year time scales.}
{Mrk~421 was found in different $\gamma$-ray emission states during the 2.3 year long observation period: The flux above 400\,GeV spans from the minimum nightly value of $(1.3\pm0.4)\cdot10^{-11}\text{cm}^{-2}\text{s}^{-1}$ to the about 24 times higher maximum flux of $(3.1\pm0.1)\cdot10^{-10}\text{cm}^{-2}\text{s}^{-1}$. Flares and different levels of variability in the $\gamma$-ray light curve could be identified with the Bayesian Block algorithm. 
The same behaviour of a quiet and active emission was found in the X-ray light curves measured by \textit{Swift}/BAT and the \textit{RXTE}/ASM, with a direct correlation in time. The behaviour of the optical light curve of GASP-WEBT and the radio light curves by OVRO and Mets\"ahovi are different as they show no coincident features with the higher energetic light curves and a less variable emission. The fractional variability is overall increasing with energy. The comparable variability in the X-ray and VHE bands and their direct correlation during both high- and low-activity periods spanning many months show that the electron populations radiating the X-ray and $\gamma$-ray photons are either the same, as expected in the Synchrotron-Self-Compton mechanism, or at least strongly correlated, as expected in electromagnetic cascades.}{}


\maketitle

\section{Introduction}
\label{sec:Introduction}

Markarian~421 (Mrk~421) is a high-frequency peaked BL Lac object (HBL) at a redshift of $z$=0.030 \citep{1999ApJ...525..176P}. It was the first extragalactic TeV emitter to be detected \citep{1992Natur.358..477P}. 

Blazars are Active Galactic Nuclei (AGN) where the jet is aligned to our line-of-sight. This means that it is possible to observe very high energy (VHE) $\gamma$-rays that are produced inside the jets and relativistically beamed in our direction. Additionally, AGN emit radiation over the whole electromagnetic spectrum, from radio wavelengths to VHE $\gamma$-rays. 

Blazars feature a spectral energy distribution (SED) with a two-bump structure. The low energy component is due to the synchrotron radiation caused by electrons of the relativistic beam, while the high energy peaked bump is attributed to other interactions. This can be the Compton scattering of less energetic photons by the same electron population in leptonic scenarios or these photons could be produced inside hadronic interactions of e.g. protons in the jet. In HBL objects as Mrk~421, the Synchrotron bump covers the energy range from radio to X-ray wavelengths while the peak can be found between UV and X-ray wavelengths. The second bump extends from low-energy $\gamma$-rays to VHE $\gamma$-rays.

A characteristic feature of blazars, and of Mrk~421 in particular, is that they show states of high activity in which the emitted electromagnetic radiation can increase by more than one order of magnitude on time scales ranging from years down to minutes. During high states blazars often show significant spectral flux changes, and up to some extent, correlated flux variations in the low- and high-energy bumps. This blazar variability is an extraordinary opportunity to break degeneracies between the various emission models. Different models produce flux variations (at a given energy band) with particles of different energies, cooling times, and cross sections for different processes, and thus are in principle distinguishable. It is also important to note that the blazar emission zone is unresolved for all instruments (with perhaps the exception of radio VLBA interferometric observations), and hence variability is the only way of probing its structure. Therefore, while ``snapshot'' multi-wavelength (MWL) spectra provide us with clues on the emission mechanisms and physical parameters inside relativistic jets, detailed studies of time variability bring us additional information on the emission mechanisms and on the structure and the dynamics of the jet itself. 

Mrk~421 has shown periods of large X-ray and $\gamma$-ray activity of various time scales, as reported previously in various publications (e.g. \citet{1996Natur.383..319G, 2004ApJ...605..662C, 2010A&A...524A..48T}).
Mrk~421 has been the target of several past MWL campaigns, with the correlation between X-rays and TeV $\gamma$-rays as one of the key features under investigation. The details in the correlation between these two bands in Mrk~421 is crucial because it relates to the energy regions where most of the power is emitted (approximately the peaks of the two SED bumps), and hence the regions of the SED which can best distinguish between different theoretical scenarios. 

A direct correlation between X-rays and TeV $\gamma$-rays has been reported multiple times during flaring activity (\citet{1995ApJ...449L..99M, 1996ApJ...472L...9B, 2004NewAR..48..419F, 2007ApJ...663..125A, 2011ApJ...734..110B, 2008ApJ...677..906F, 2009ApJ...691L..13D, 2011ApJ...736..131A, 2011ApJ...738...25A, 2013PASJ...65..109C, 2015A&A...578A..22A}). Recently, \citet{2015A&A...576A.126A} and \citet{2016ApJ...819..156B} also reported the existence of this correlation during low activity.
\citet{2005ApJ...630..130B, 2009ApJ...695..596H} were able to constrain the correlation to time differences below 1.5 days. These results are in agreement with the Synchrotron-Self-Compton (SSC) model, where the photons from both X-ray and $\gamma$-ray energies are produced by the same electron population. Other authors reported orphan flares in TeV $\gamma$-rays without an X-ray counterpart, which were observed in Mrk~421 during a MWL campaign in 2003 and 2004 \citep{2005ApJ...630..130B}, unable to be explained by the SSC model. In \citet{2009ApJ...703..169A} a correlation between TeV and X-rays is not found, and a possible hadronic origin of the emission is discussed. However, this correlation study relates to short observations (two half-day long observations) with very low variability. The X-ray emission was accurately characterized with continuous XMM observations, and flux variations at the level of 10\% could be significantly resolved. Yet the TeV $\gamma$-ray measurements covered only a small fraction of the XMM observations, and had relatively large error bars. Therefore, the presented X-ray/TeV correlation results in \citet{2009ApJ...703..169A} were not conclusive, and show very clearly the importance of having long, well sampled and sensitive TeV $\gamma$-ray observations to perform this kind of studies.\\

Other energy bands are not evidently correlated with X-rays and TeV $\gamma$-rays. \citet{1995ApJ...449L..99M, 2007ApJ...663..125A, 2013PASJ...65..109C} report about a missing correlation of the optical and UV emission to the X-ray and TeV $\gamma$-ray emission. Confirming the trend of a strong correlation between X-rays and VHE, the work of \citet{2016ApJ...819..156B} also reports a lack of correlation between optical/UV and X-rays, and moreover ascribes the observed broadband variability features during low activity to in situ electron acceleration in multiple compact regions. In \citet{2009ApJ...695..596H} a correlation with a time lag between the optical and the TeV $\gamma$-ray light curves is found, once with the optical features leading the TeV features and once vice versa, but the likelihood to have observed the optical leading and lagging the TeV features by chance is 20\% and 60\% respectively. In \citet{2015A&A...576A.126A} an anti-correlation between the optical and UV light curves with the X-ray light curves is reported, but with the possibility that might have been found by chance, proposing a dedicated correlation analysis over many years in order to properly characterize the temporal evolution of the optical and X-ray/TeV $\gamma$-ray bands.\\
An evidence of a correlation between radio and $\gamma$-ray activity was reported in \citet{2003A&A...410..101K}, where the study of a single radio outburst with a X-ray and TeV $\gamma$-ray counterpart in February-March 2001 is presented. The author models a scenario in which the acceleration of electrons in the middle part of the jet describes well the temporal evolution of such a multispectral flare.\\
In the more recent work of \citet{2014A&A...571A..54L} a marginally significant correlation between radio and GeV $\gamma$-rays (without time lag) is reported. This study used observations spanning many months from 2011, when Mrk~421 did not show any flaring activity, hence suggesting a co-location of the radio and $\gamma$-ray emission of Mrk~421 during typical (low) activity.\\

A different result is derived from the outstanding radio activity observed in September 2012, where Mrk~421 showed a particularly symmetric flare profile, with the highest radio flux measured in three decades, as reported in \citet{2014MNRAS.445..428M} and \citet{2015MNRAS.448.3121H}. Both works assume that this giant radio flare is physically connected to a large $\gamma$-ray flaring activity measured by \textit{Fermi}-LAT about one month before, and \citet{2014MNRAS.445..428M} uses this time difference to locate the origin of the $\gamma$-ray emission upstream of the radio emission.\\
 
Because of the above-mentioned complexity and sometimes controversy in the multi-band flux variations and correlations observed during relatively short (weeks to months) campaigns, one needs very long (multi-year) campaigns in order to put things into context. In this paper we report an extensive study of the multi-band flux variability of Mrk~421 during the 2.3 year long period that spans from February 2007 to July 2009. We adopted the methodology reported in \citet{2015A&A...576A.126A}, which had been applied in a much shorter multi-instrument data set.

There are several publications that report studies with the VHE $\gamma$-ray emission of Mrk~421 during the above mentioned 2.3 year long period; yet they typically relate to smaller temporal intervals. For instance, \citet{2012A&A...542A.100A} reported MAGIC observations of a high active state performed from December 2007 to June 2008, and \citet{2011ApJ...736..131A} and \citet{2015A&A...576A.126A} reported results related to observations from a 4.5 months long time interval from January to June 2009. A very interesting study using Whipple~10m observations performed from December 1995 to May 2009 was reported in \citet{2014APh....54....1A}, which allowed to study the duty cycle and to evaluate the VHE emission and its correlation with the X-ray emission. The study that we report in this paper relates to a time period that is (almost) contained in \citet{2014APh....54....1A}, but it provides a large number of improvements such as the larger sensitivity of MAGIC with respect to Whipple~10m, which allows to resolve the VHE flux with smaller uncertainties, and hence to study the variability and its correlation on shorter time scales (2-days). Moreover, in this paper we apply a more sophisticated treatment to quantify variability and correlations (adopted from \citet{2015A&A...576A.126A}), and we extend the study to extensive light curves collected at radio, optical and hard X-rays (above 15 keV), hence overall giving a more complete picture of the year-long multi-band flux variability of Mrk~421 than the one given in \citet{2014APh....54....1A}.\\

The paper is organised as follows. Section~\ref{sec:MAGICMrk421} describes the MAGIC observations, as well as the analysis and results obtained. Section~\ref{sec:Bayesian} describes the application of the Bayesian Block algorithm to the MAGIC data, and the resulting quantification of the flux variability and identification of several VHE flares. The Bayesian block is a well established methodology, but this is the first time that it is applied to VHE data. Section~\ref{sec:other} describes the extensive observations of Mrk~421 performed at radio, optical and X-rays, and in Sections~\ref{sec:Variability} and \ref{sec:Correlations} we report the quantification of the multi-band variability and its correlations. Finally, in Section~\ref{sec:Summary} and \ref{sec:Discussion} we summarise and discuss the results presented.

\section{MAGIC observations of Mrk~421}
\label{sec:MAGICMrk421}
\subsection{The MAGIC telescopes}
\label{subsec:MAGIC}
The MAGIC (Major Atmospheric Gamma-ray Imaging Cherenkov) telescopes are a system of two Cherenkov telescopes with a mirror diameter of 17\,m each. They are situated at the ORM (Observatory Roque de los Muchachos) on the Canary Island of La Palma at a height of 2200\,m above sea level.

In 2004 the MAGIC-I telescope was commissioned and started its observations in single telescope mode. The performance during the stand-alone operation of MAGIC-I was presented in \citet{2008ApJ...674.1037A} and \citet{2009APh....30..293A}. Stereoscopic data were taken after the second telescope, MAGIC-II, was commissioned in 2009, and a major upgrade of the MAGIC telescopes was performed in 2012 \citep{2016APh....72...61A,2016APh....72...76A}. 

\subsection{Observations and data analysis}
\label{subsec:observations}
Mrk~421, one of the strongest and brightest extragalactic sources, is observed by MAGIC on a regular basis. The source is observable from late November to June from the MAGIC latitude. In this analysis data of Mrk~421 of MAGIC-I in single telescope operation from 8th March 2007 (MJD~54167) to 15th June 2009 (MJD~54997), a time span of over two years, were examined. The overall amount of good quality data taken in Wobble mode \citep{1994APh.....2..137F} are 95.6 hours distributed over 95 observation nights. The data cover a zenith angle range from 9$^\circ$ to 45$^\circ$. Data with too bright sky conditions and bad weather conditions were excluded. The data analysis was carried out using the standard MAGIC analysis chain MARS (MAGIC Analysis and Reconstruction Software) \citep{Zanin2013}. During the selected time span an integral sensitivity as low as 1.6\% of the Crab Nebula flux is reached and the energy resolution is $\sim$ 20\% \citep{2009APh....30..293A}.

\subsection{Measured VHE $\gamma$-ray flux}
\label{subsec:MAGICresults}

The light curve of Mrk~421 measured by MAGIC-I is binned nightly and is shown in both Figure~\ref{fig:BB} and in the top panel of Figure~\ref{fig:5lightcurves}. 

The light curve is naturally divided into three observation cycles due to the observability gaps of Mrk~421 from the end of June to the end of November each year with the MAGIC telescopes. The three time periods will be called Period~1 for data from February 2007 to August 2007, Period~2 for data from September 2007 to beginning of September 2008, and Period~3 for data from beginning of September 2008 to July 2009. In Figure~\ref{fig:5lightcurves} these three periods are marked. The light curve shows different levels of source flux and variability in these three time spans. In Period~1 and in Period~3 the flux is clearly at a lower level than in Period~2. \\

During Period~1 the average flux of the six data points is at a level of \mbox{(0.38$\pm$0.03)\,CU}\footnote{A Crab Unit is defined here as a flux of $8.08 \cdot 10^{-11}$\,cm$^{-2}$s$^{-1}$ in the energy range from 400\,GeV to 50\,TeV \citep{2008ApJ...674.1037A}.}. The flux is variable with variations up to a factor of 2 around the average flux. During Period~2 the flux is at a high average level of \mbox{(1.38$\pm$0.02)\,CU} and it seldomly falls below 1\,CU. The light curve shows a high variability with flux variations of about a factor of 3 around the average. The flux varies between the lowest value of \mbox{(0.4$\pm$0.1)\,CU} on 17th December 2007 and the maximum value of \mbox{(3.8$\pm$0.1)\,CU} on 31st March 2008 (MJD~54556). During Period~3 the average VHE $\gamma$-ray flux is \mbox{(0.61$\pm$0.01)\,CU} with variations of up to a factor of $\sim$ 2.\\

The time-averaged fluxes detected by MAGIC for the three identified observation periods are comparable to the ones measured by the Whipple~10m telescope for the seasons 2006-2007, 2007-2008, and 2008-2009 respectively, which were reported in \citet{2014APh....54....1A}. The Whipple telescope detected a flux of \mbox{(0.28$\pm$0.02)\,CU} for the 2006-2007 data, which is at a comparable level with the \mbox{(0.38$\pm$0.03)\,CU} for Period~1 of the MAGIC data (here it has to be noted that the Whipple observations cover a larger time span, which already starts in 2006). Then an average flux of \mbox{(1.46$\pm$0.09)\,CU} is reported by Whipple in 2007-2008, similar to the value of \mbox{(1.38$\pm$0.02)\,CU} for Period~2, confirming the higher flux state. For the 2008-2009 season, the Whipple flux was \mbox{(0.55$\pm$0.03)\,CU}, which is comparable to the average flux of \mbox{(0.61$\pm$0.01)\,CU} in Period~3 measured by MAGIC.

In summary, during the three observation periods covered in this paper, Mrk~421 showed three clearly distinct VHE flux levels, with different apparent levels of variability. A quantitative evaluation of the VHE flux variability in these three periods is reported in sections 3 and 5, following the prescriptions given in \citet{1998ApJ...504..405S}, \citet{2013ApJ...764..167S} and \citet{2015A&A...573A..50A}.

\section{Bayesian Blocks}
\label{sec:Bayesian}

We applied the Bayesian Block algorithm \citep{1998ApJ...504..405S, 2013ApJ...764..167S} to the TeV light curve of Mrk~421. The algorithm generates a block-wise constant representation of a sequential data series by identifying statistically significant variations, and is suitable to characterize local variability in astronomical light curves, even when not evenly sampled.

The optimal segmentation (defined by its change points) maximizes the goodness-of-fit with a certain model for the data lying in a block. The method requires a prior probability distribution parameter (ncp$_{\text{prior}}$) for the number of changing points (N$_{\text{cp}}$), a kind of smoothing parameter derived from the assumption that $\text{N}_{\text{cp}} \ll \text{N}$, the number of measurements. A false-positive rate (p$_0$) is associated to the choice ncp$_{\text{prior}}$.

\begin{figure*}[tb!]
\centering
    \includegraphics[width=0.99\textwidth]{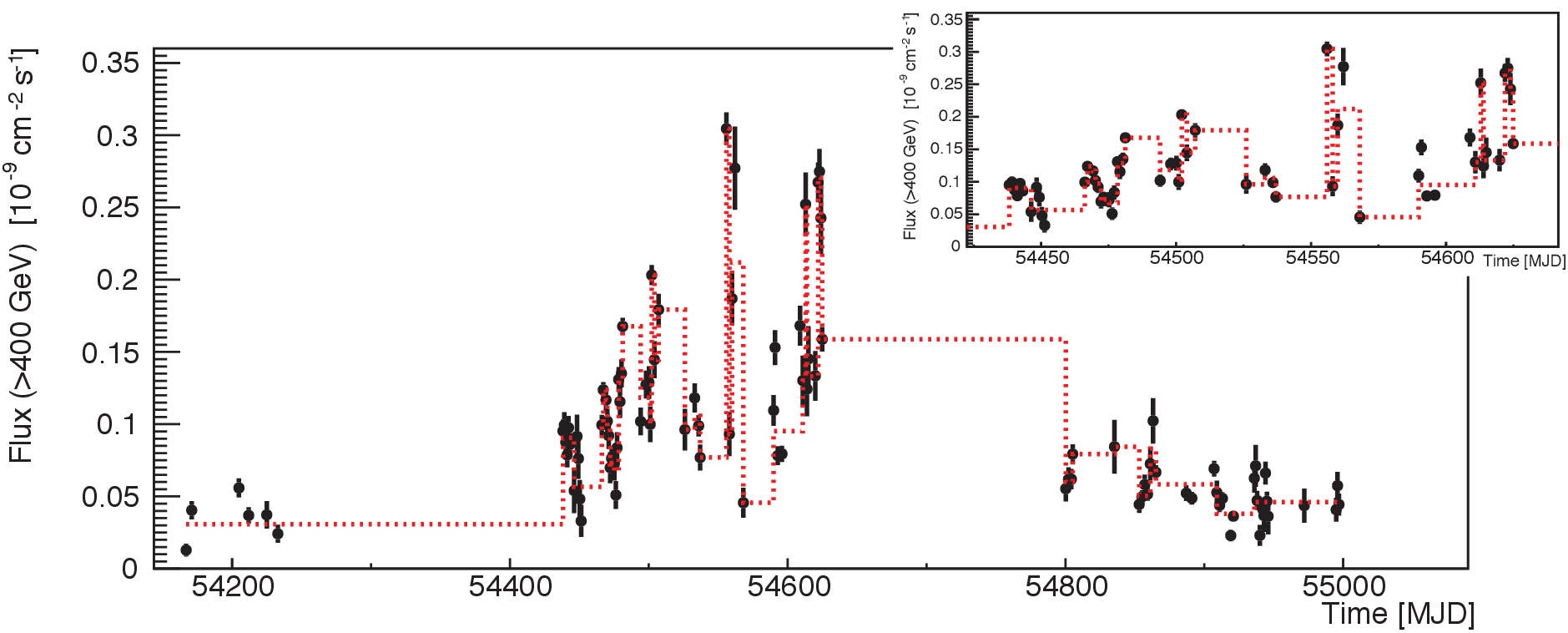}  
    \caption{Bayesian Blocks representation of the MAGIC light curve (black dots) from March 2007 to June 2009. The red dotted line defines the different identified blocks. The inlay shows a zoomed version for the time range from December 2007 to June 2008, the high active Period~2. The long flat lines with no sampling between a data point and a new block do not guarantee a stable flux.}
    \label{fig:BB}
\end{figure*}

The false-positive rate was chosen to be p$_0$=0.01, leading to a ncp$_{\text{prior}}$=3.92. We obtained the 39 blocks representation for 95 data points shown as a red dotted line in Figure \ref{fig:BB} on top of the flux points measured by MAGIC (black dots). The height of each block is the weighted average of all integral fluxes belonging to it.

An advantage of the Bayesian Block algoritm is that it is able to identify significant changes in data series independently of variations in gaps or exposure. Therefore, no information on true or important flux changes is lost, as it can happen when applying other techniques where the data series is binned in predefined temporal intervals.

This is the first time that the Bayesian Block algorithm was applied to a VHE $\gamma$-ray light curve. We use the results to estimate the variability level of the light curve in the different observation periods and to define flares. To quantify the variability for each period, we can simply determine the ratio of resulting number of blocks and the number of data points. A higher ratio implies a higher flux variability. All six data points from Period~1 belong to the same initial block. This ratio of 1/6 indicates a low variability during this period. The lack of additional blocks during this period may also be related to the very low number of data points. The high activity in Period~2 is evident by the 30 blocks detected for 56 data points during this time period by the algorithm (see inlay of Figure~\ref{fig:BB}). The resulting ratio of 30/56, which is slightly above 0.5, shows that the light curve is substantially more variable than Period~1. In Period~3 we have an 8-block representation for 33 data points, which is a ratio of $\sim$0.24. This lower variability of the light curve during this period shows a milder activity of the AGN than in Period~2. An additional discussion on the variability will be given in Section~\ref{sec:Variability}.\\

It is of great interest to identify flaring activities in light curves, but the definition of a flare is somewhat arbitrary and, since blazars vary on time scales from years down to minutes, a definition is strongly biased by the prejudice of the temporal bins used to produce the light curves. It is easy to miss flaring activities in light curves with ``too large'' temporal bins (if the variability occurs on small time scales) or in light curves with ``too small'' temporal bins (if the flux values are dominated by statistical uncertainties). In this context, the Bayesian Block algorithm benefits from a more suitable temporal split (according to the true variability), and hence it can be used as a very efficient method to find flares. In the following, VHE $\gamma$-ray flares are defined as a flux rise of at least a factor of 2. This comparison is based on the block heights, i.e. the weighted average flux of all data points in one block. A flare can include several rising steps in a row, which add up to a local maximum in flux. Subsequently, the flux decreases to a lower flux, which can happen on a daily or longer time scale. By using this flux-doubling threshold we could identify several flares, which are reported in Table~\ref{tab:flares}. \\

We estimate the flux-doubling times using the height difference between consecutive blocks and the time between the last data point of a given block and the starting point of the next block, which is a conservative measure of the rise time between blocks. In the case of several consecutive flux rises among continuous blocks, the flux-doubling time reported in Table~\ref{tab:flares} considers the rise as a single increase from minimum to maximum. It can be seen that the flux doubles its value on different time scales. The flux-doubling can occur during just one night, e.g. for the block starting on MJD~54502, but it can also take many days. Additionally, it should be noted that it cannot be ruled out that the flux might fall between two measurements. All determined flux-doubling times are subject to this possibility. For the first entry in the table the flux-doubling time of 139 days is not meaningful because the time interval includes the long observation gap from May to December 2007 where the source behaviour in $\gamma$-rays is unknown. Therefore, the flux-doubling times reported in Table~\ref{tab:flares} should be considered as upper limits to the actual time needed to double the flux. That is, the actual flux-doubling times could be shorter than the ones reported.\\

\begin{table*}[tb!]
\caption{Dates, factor of flux increase and flux-doubling times of flares found by the Bayesian Block algorithm for the MAGIC light curve. The given MJD identifies the first day of the highest block. See definition of VHE flare in the text.}
\label{tab:flares}
\centering
\begin{tabular}{p{1cm}p{1.5cm}p{2.2cm}p{12.0cm}}
\toprule
\textbf{MJD}	&\textbf{increase}	&\textbf{flux-doubling time [days]}	&\textbf{notes}\\
\midrule
\midrule
54438 & $3.0\pm 0.3$ & 139* & The flux rise follows the low flux in the beginning of 2007.\\
\midrule
54467 & $2.2\pm 0.3$ & 15 & Two subsequent rises.\\ 
\midrule
54481 & $2.5\pm 0.3$ & 3 & Two subsequent rises. \\
\midrule
54502 & $2.0\pm 0.3$ & 1 & Flux rise in just one night.\\
\midrule
54556 & $4.0\pm 0.6$ & 10 & Rise to the overall maximum flux value. The last given data point before this block was taken 19 days before. \\
\midrule
54560 & $2.3\pm 0.5$ & 2 & \\
\midrule
54613 & $5.5\pm 1.7$ & 20 & The rise to the maximum takes place in three single steps. The first rise of a factor of 2.1 follows an observation 21 days before. The following block has a length of 19 days. Subsequently, the flux rises by a factor of 1.4 in just two days and by a factor of 1.9 during the same time interval.\\
\midrule
54622 & $2.0\pm 0.2$ & 2 & \\
\bottomrule
\end{tabular}
\begin{flushleft}
*: Includes an observation gap of about half a year.
\end{flushleft}
\end{table*}

\begin{figure*}[t]
\centering
    \includegraphics[width=0.98\textwidth]{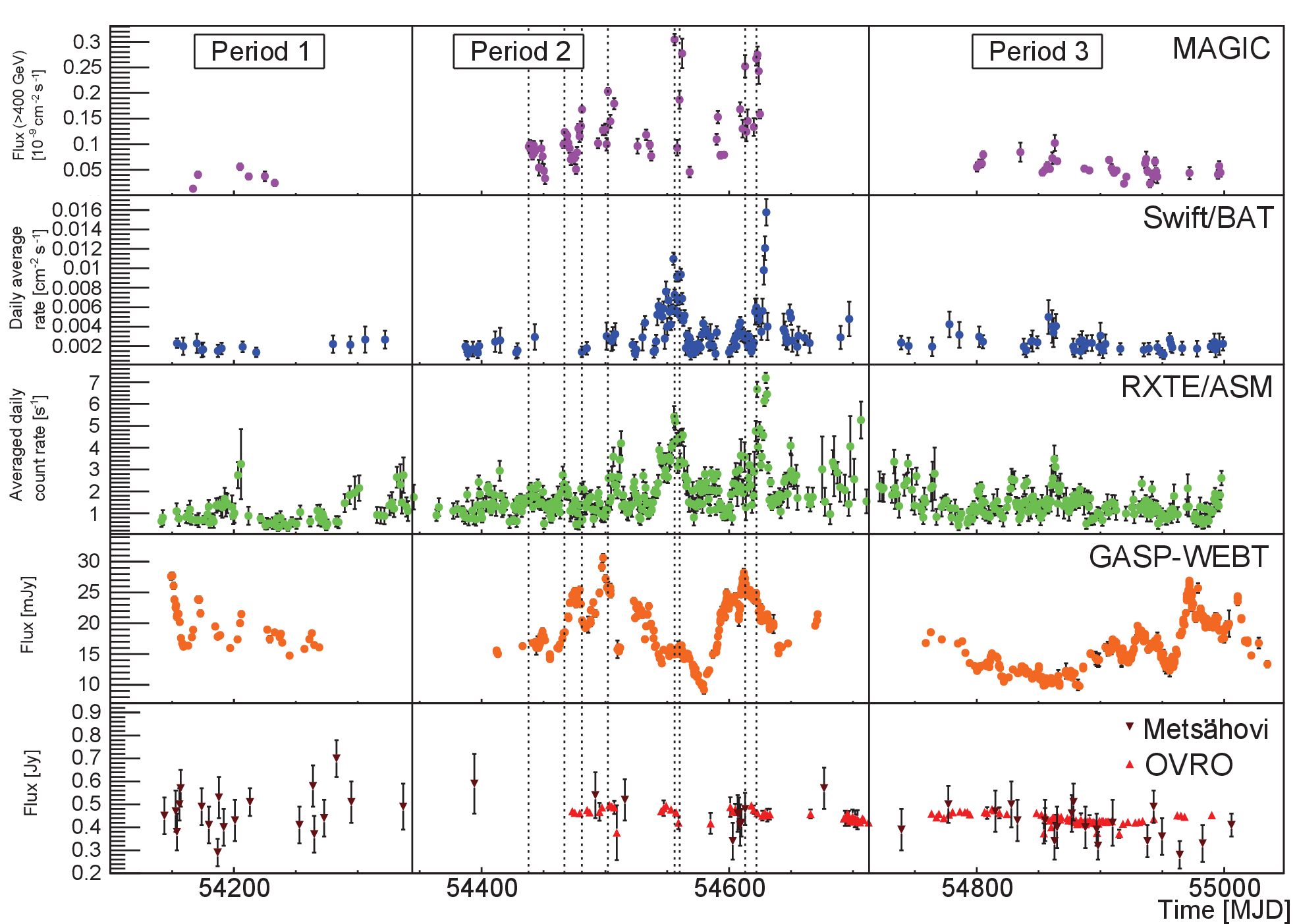}  
    \caption{Light curves of MAGIC, \textit{Swift}/BAT, \textit{RXTE}/ASM, GASP-WEBT, Mets\"ahovi and OVRO from top to bottom in the time range from February 2007 to July 2009. The vertical dotted black lines denote the position of the TeV $\gamma$-ray flares as identified with the Bayesian Block algorithm (see Section~\ref{sec:Bayesian}). The vertical black lines mark the division between the three time periods (Period~1, Period~2, Period~3).} 
    \label{fig:5lightcurves}
\end{figure*}

The flares identified by using the Bayesian Block algorithm are marked in Figure~\ref{fig:5lightcurves} by vertical dotted lines so that it is possible to compare these positions with features in the light curves in the other wavelengths.

\section{Observations at X-ray, optical and radio wavelengths}
\label{sec:other}

To study the variability and correlation between the TeV $\gamma$-ray data and other wavebands, data from several other instruments were considered.  In the X-ray range data from \textit{Swift}/BAT and \textit{RXTE}/ASM were selected. The optical data shown here is from the GASP-WEBT consortium (which includes data from the KVA telescope located at the ORM close to MAGIC). Data from the Mets\"ahovi and OVRO telescopes are used in the radio range.

\subsection{Hard X-ray observations with \textit{Swift}/BAT}
\label{subsec:Swift}
The Burst Alert Telescope (BAT) on board the \textit{Swift} satellite observes Mrk~421 in the hard X-ray regime, from 15 to 50\,keV \citep{2013ApJS..209...14K}. The \textit{Swift}/BAT transient monitor results are provided by the \textit{Swift}/BAT team\footnote{http://swift.gsfc.nasa.gov/results/transients/weak/Mrk421.lc.txt}. Considering only averaged daily rates with a rate to rate error ratio greater than two, and additionally discarding six measurements with negative rates (on MJD~54288, 54476, 54638, 54750, 54914, and 54981), results in a total of 821 hours of data distributed over 168 nightly flux measurements between 23rd February 2007 (MJD~54154) and 17th June 2009 (MJD~54999).

The \textit{Swift}/BAT light curve of Mrk~421 is shown in Figure \ref{fig:5lightcurves}. The overall hard X-ray flux behaviour is comparable to that of the MAGIC light curve, with a higher activity in Period~2 and several features that appear to be coincident, like the peak structure around MJD~54560.

\subsection{Soft X-ray observations with \textit{RXTE}/ASM}
\label{subsec:RXTE}
The All-Sky Monitor (ASM) was an instrument on board the \textit{RXTE} satellite. It observed Mrk~421 in the energy range from 2 to 10\,keV \citep{1996ApJ...469L..33L}.

The results shown here are provided by the ASM/\textit{RXTE} teams at MIT and at the \textit{RXTE} SOF and GOF at NASA's GSFC\footnote{xte.mit.edu/asmlc/ASM.html}. Only averaged daily count rates, each consisting of several so-called observation dwells of 90\,s length, with a rate to rate error ratio greater than two are considered for the following studies. Additionally, two negative rates, on MJD~54371 and 54914, are discarded. This results in a total of 532 daily flux measurements with a total observation length of 260 hours between 10th February 2007 (MJD~54141) and 16th June 2009 (MJD~54998).

The \textit{RXTE}/ASM light curve of Mrk~421 is shown in Figure \ref{fig:5lightcurves}. The soft X-ray flux shows a similar behaviour to that of the hard X-rays and VHE $\gamma$-rays, which includes several overall flux levels and peak structures that are present also in the \textit{Swift}/BAT and MAGIC light curves. 

\subsection{Optical observations}
\label{subsec:KVA}
The optical data in the R-Band shown here were recorded by the KVA (Kungliga Vetenskapsakademien) telescope and a collection of telescopes, which work together in the GASP-WEBT (Whole Earth Blazar Telescope)\footnote{http://www.oato.inaf.it/blazars/webt/} consortium \citep{2008A&A...481L..79V}. The KVA telescope is situated at the ORM on La Palma close to the MAGIC telescopes. Photometric observations in the R-Band are made with a 35\,cm telescope. Observations are carried out in the same time intervals as MAGIC observations. Optical observations of Mrk~421 by the KVA telescope started in 2002, and show a variable optical light curve \citep{2008AIPC.1085..705T}.

Mrk~421 is regularly monitored by telecopes of GASP-WEBT, and KVA in particular. The optical data reported in this paper relate to the period from 18th February 2007 (MJD~54149) to 23rd July 2009 (MJD~55035), which were recorded by the following instruments: Abastumani, Castelgrande, Crimean, L'Ampolla, Lulin, KVA, New Mexico Skies (now called iTelescopes), Sabadell, St. Petersburg, Talmassons, Torino, and Tuorla observatories. It should be mentioned that the flux measurements are corrected for the contribution of the host galaxy (see \citet{2007A&A...475..199N}) as well as for galactic extinction \citep{2011ApJ...737..103S}.

The GASP-WEBT light curve shown in Figure~\ref{fig:5lightcurves} includes a total of 815 observations distributed over 353 nights. 
When comparing the optical light curve to the $\gamma$-ray and X-ray light curves it is important to note that the optical light curve cannot be separated into different activity phases as the other light curves. The flux varies by the same amount throughout the whole observation length of more than two years. It can be seen that the features in the GASP-WEBT light curve are longer than and not coincident with those of the MAGIC, \textit{RXTE}/ASM and \textit{Swift}/BAT light curves.

\subsection{Radio observations with Mets\"ahovi}
\label{subsec:Metsahovi}

Radio data at 37\,GHz are recorded by the 13.7\,m telescope at the Mets\"ahovi Radio Observatory in Finland. \citep{1998A&AS..132..305T}

Considering only data points with a flux to error ratio greater than four of the Mets\"ahovi light curve, leaves 49 nightly flux measurements between 13th February 2007 (MJD~54144) and 24th June 2009 (MJD~55006). The light curve is shown in Figure \ref{fig:5lightcurves}. In comparison to the VHE $\gamma$-ray, the X-ray and the optical light curves mentioned above, the overall radio flux measured by Mets\"ahovi is rather stable, yet with a slight decrease in Period~3.

\subsection{Radio observations with OVRO}
\label{subsec:OVRO}
The Owens Valley Radio Observatory, located in the USA, operates a 40\,m radio telescope measuring at 15\,GHz. It started observations in January 2008 and therefore does not cover the whole time span of MAGIC observations\footnote{www.astro.caltech.edu/ovroblazars/data/data.php}. \citep{2011ApJS..194...29R}

In the available data set, often two observations were made during one day, which were only separated by $\sim$ 2 minutes. These data points were averaged, which results in a total of 119 data points. The light curve with data points between 8th January 2008 (MJD~54473) and 8th June 2009 (MJD~54990) is shown in Figure \ref{fig:5lightcurves}. As it occurs with the Mets\"ahovi light curve, the flux is rather stable, with a small decrease in Period~3.

\section{Multi-band flux Variability}
\label{sec:Variability}

In order to quantify the variability in the emission of Mrk~421, the fractional variability $F_\text{var}$, as it is given in equation 10 in \citet{2003MNRAS.345.1271V}, is used. It is calculated using

\begin{equation}
F_\text{var} = \sqrt{\dfrac{S^2 - \overline{\sigma^2_\text{err}}}{\overline{x}^2}},
\end{equation}

and represents the normalized excess variance. $S$ is the standard deviation and $\overline{\sigma^2_\text{err}}$ the mean square error of the flux measurements. $\bar{x}$ stands for the average flux. The uncertainty of $F_\text{var}$ is given by equation 2 in \citet{2015A&A...573A..50A}, after \citet{2008MNRAS.389.1427P}:
\begin{equation}
\Delta F_\text{var} = \sqrt{F^2_\text{var}+\text{err}(\sigma^2_\text{NXS})} - F_\text{var},
\end{equation}
where $\text{err}(\sigma^2_\text{NXS})$ is given by equation 11 of \citet{2003MNRAS.345.1271V}:
\begin{equation}
\text{err}(\sigma^2_\text{NXS}) = \sqrt{\left(\sqrt{\dfrac{2}{N}}\cdot\dfrac{\overline{\sigma^2_\text{err}}}{\bar{x}^2}\right)^2+\left(\sqrt{\dfrac{\overline{\sigma^2_\text{err}}}{N}}\cdot\dfrac{2F_\text{var}}{\overline{x}}\right)^2} .
\end{equation}

Here, $N$ is the number of data points in a light curve. Note from equation~1 that $F_\text{var}$ is not defined (and hence cannot be used) when the excess variance is negative, which can occur in the absence of variability, or when the instrument sensitivity is not good enough to detect it (i.e. large flux uncertainties).

\begin{figure}[tb]
\centering
    \includegraphics[width=0.49\textwidth]{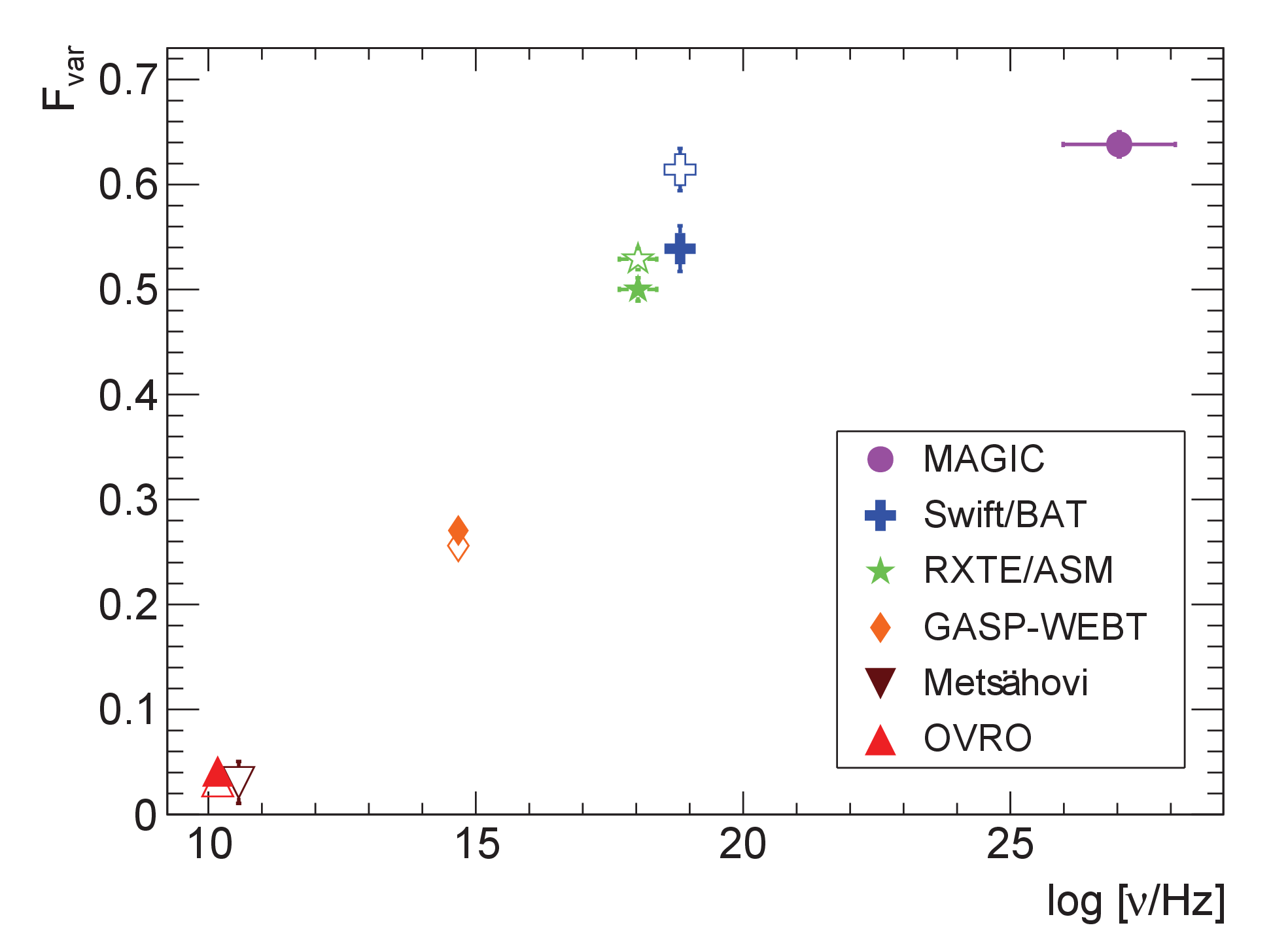}  
    \caption{Fractional variability ($F_\text{var}$) as a function of the frequency for the 2.3 year long time range from February 2007 to July 2009. The fractional variability was computed in two different ways: using all the flux measurements from the light curves reported in Figure~\ref{fig:5lightcurves} (depicted with open markers), and using only those observations simultaneous to the VHE $\gamma$-ray measurements from MAGIC (depicted with filled markers). Vertical bars denote 1$\sigma$ uncertainties and horizontal bars indicate the width of each energy bin.}
    \label{fig:fvar}
\end{figure}

$F_\text{var}$ is calculated for all the light curves shown in Figure~\ref{fig:5lightcurves} and the results are shown in Figure~\ref{fig:fvar} with open markers. For MAGIC, \textit{Swift}/BAT, \textit{RXTE}/ASM, Mets\"ahovi and OVRO, the shown light curves feature one data point per night. For GASP-WEBT, the light curve contains nights with more than one data point. For the calculation of $F_\text{var}$, the multiple GASP-WEBT optical fluxes related to single days were averaged, thus obtaining a single value.

In order to improve the direct comparison of the variability determined for the various energy bands, we also computed $F_\text{var}$ using only the multi-instrument observations strictly simultaneous to those performed by MAGIC. These $F_\text{var}$ values are depicted by the filled markers in Figure~\ref{fig:fvar}, and remove potential biases due to the somewhat different temporal coverage of the various instruments.

The overall behaviour of the fractional variability shows a rising tendency with increasing frequency. Considering only the $F_\text{var}$ values determined with simultaneous multi-instrument observations (filled markers in Figure~\ref{fig:fvar}), the highest variability occurs in the VHE $\gamma$-ray band measured by MAGIC ($F_\text{var}=0.64\pm0.01$), although it is quite similar to the variability measured in the soft X-ray band ($F_\text{var}=0.50\pm0.01$) and hard X-ray band ($F_\text{var}=0.54+\pm0.02$) by \textit{RXTE}/ASM and \textit{Swift}/BAT respectively.\\

\begin{figure}[tb]
\centering
    \includegraphics[width=0.49\textwidth]{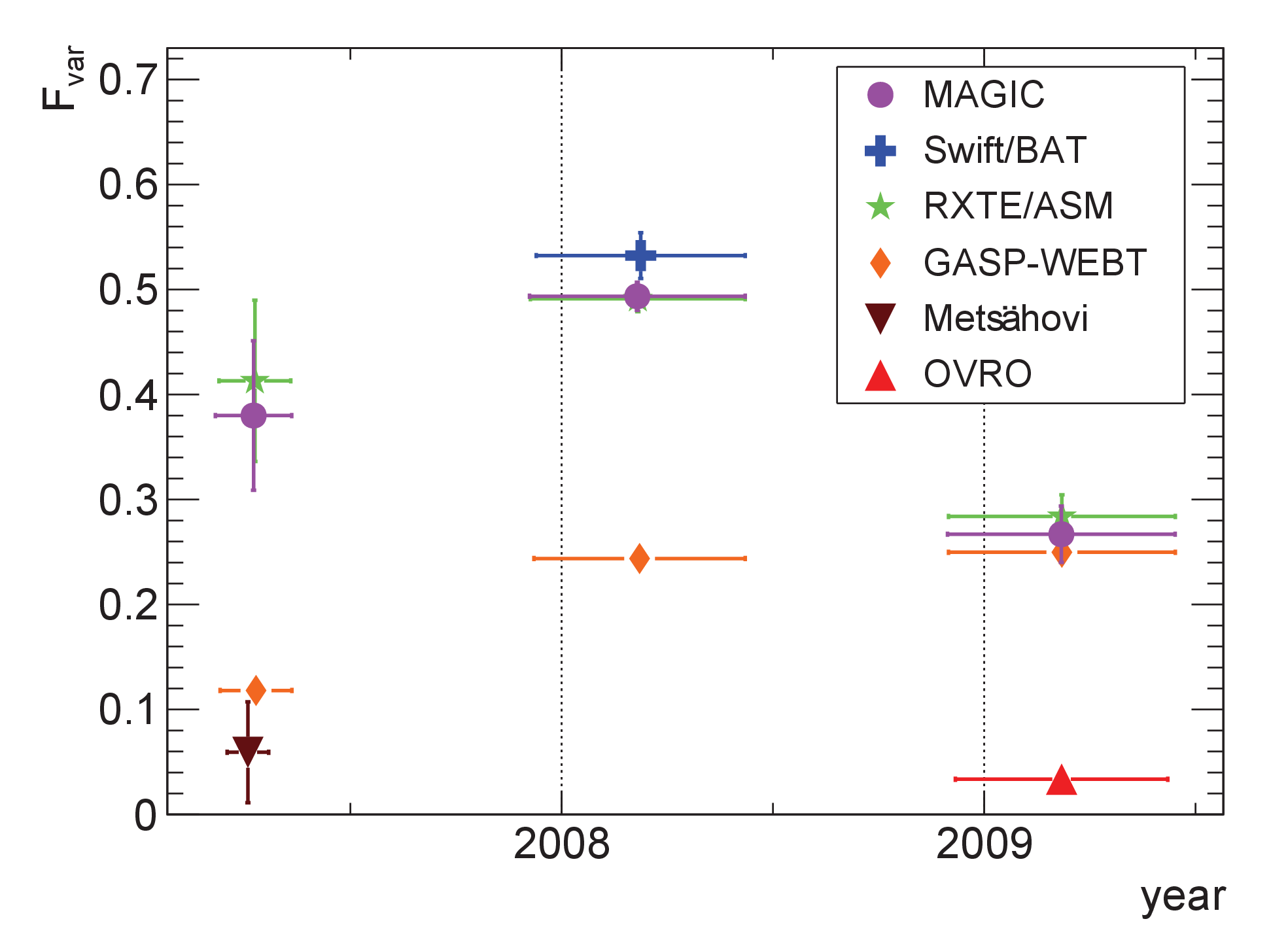}  
    \caption{Multi-instrument fractional variability ($F_\text{var}$) for the three periods defined in Figure~\ref{fig:5lightcurves}. The fractional variability was computed using only those observations simultaneous to the VHE $\gamma$-ray measurements from MAGIC. Vertical bars denote 1$\sigma$ uncertainties and horizontal bars indicate the covered time span of each instrument.}
    \label{fig:fvarsingleyears}
\end{figure}

As mentioned in the previous sections (e.g. see Figure~\ref{fig:5lightcurves}), the overall flux levels and source activity appear different for the three different observation periods. Figure~\ref{fig:fvarsingleyears} reports the multi-band fractional variability determined separately for Periods~1, 2 and 3. The main trend observed in the 2.3 year long time span reported in Figure~\ref{fig:fvar} is also reproduced when splitting the data in the three different periods: $F_\text{var}$ always increases with energy, with the highest variability occurring in the X-ray and VHE $\gamma$-ray bands. The \textit{Swift}/BAT light curve with one-day temporal bins reported in Figure~\ref{fig:5lightcurves} has large statistical uncertainties, which, because of the relatively low activity and low variability of Mrk~421 during Periods~1 and 3, yielded a negative excess variance, hence preventing the calculation of the fractional variability for these two periods. On the other hand, the \textit{RXTE}/ASM light curve with one-day temporal bins reported in Figure~\ref{fig:5lightcurves} have somewhat smaller uncertainties and a better temporal coverage than that of \textit{Swift}/BAT, which permitted the quantification of the fractional variability in the soft X-ray energy band for the three temporal periods considered. 

In Figure~\ref{fig:fvarsingleyears} it can also be seen that the variability for the MAGIC light curve is higher for Period~2 than in Period~1 and 3 as it was already shown by the quantification of the variability with the results of the Bayesian Block algorithm (see Section~\ref{sec:Bayesian}). Due to the lower average flux in Period~1 compared to Period~3, the fractional variability in Period~1 is higher than in Period~3.

It is worth noticing that the fractional variability in the optical band is comparable to that at X-rays and VHE $\gamma$-rays during Period~3, which did not happen during Periods~1 and 2. Inspecting the light curves reported in Figure~\ref{fig:5lightcurves}, one can see that the time scales involved in the reported variabilities are very different. While the X-ray and VHE $\gamma$-ray light curves show day-long flux variations on the top of a rather stable flux level, the optical flux shows many-day-long flux variations on the top of a flux level that increases by about a factor of two throughout Period~3. Therefore, despite the very comparable $F_\text{var}$ values during Period~3, the emission in the optical band is probably not related to that in the X-ray and VHE $\gamma$-ray bands.

These results are consistent with results from previous publications. This includes the rising fractional variability of Mrk~421 from optical to X-ray energies in 2001 \citep{2007A&A...462...29G} and the same increase from optical to X-ray energies in March 2010 during a flare with a comparable variability of the VHE and the X-ray light curves \citep{2015A&A...578A..22A}. These results are complemented by \citet{2015A&A...576A.126A} and \citet{2016ApJ...819..156B}, which presented multi-wavelength data during the relatively low activity observed from January to June 2009 and from January to March 2013 respectively. These include results from the \textit{Fermi}-LAT closing the gap between the X-ray and TeV $\gamma$-ray energy bands. They report a low flux in radio energies, rising to a maximum in the X-ray energy band. For GeV $\gamma$-rays measured by the \textit{Fermi}-LAT the variability drops to a level comparable to the optical and UV wave band. The variability in the TeV $\gamma$-ray light curves increases to a level comparable to X-rays, which is consistent with the result from this study, that uses a much larger time span.

\section{Multi-band correlations}
\label{sec:Correlations}

To quantify the correlation of two light curves, the Discrete Correlation Function (DCF), which was introduced by \citet{1988ApJ...333..646E}, is used here. A study of the correlations of the MAGIC light curve with light curves of other wavelengths has already been done for Mrk~421 for the first half of 2009 in \cite{2015A&A...576A.126A}. In that publication a method to determine confidence intervals for the resulting correlation has been described in detail. Here, a short introduction to the method will be given. For more detailed information on that method the reader is referred to the cited publication and references therein.

The errors of the DCF values as stated by \citet{1988ApJ...333..646E} might not be appropriate when the individual light-curve data points are correlated red-noise\footnote{Red noise data is characterised by a power spectral density per unit of bandwidth proportional to 1/f$^2$, where f is the frequency. \citep{2012ApJ...749..191C}.} data \citep{2003ApJ...584L..53U}. Since this is not the case for the given light curves, a Monte Carlo based approach is applied here to determine confidence intervals for the DCF values. Therefore, 1000 light curves are simulated for each telescope which feature the same sampling pattern and comparable exposure times as the original light curve. In addition, the power spectral density (PSD) should be as similar as possible to the PSD of the original light curve. Therefore, the light curves are simulated with PSDs following a power law with spectral indices in a range from -1.0 to -2.9 in steps of 0.1. The light curves with the PSD, which match the PSD of the original light curve best, are determined using the PSRESP method \citep{2008ApJ...689...79C}. 

The DCF itself is calculated for sets of original light curves. With the calculated DCF of 1000 simulated light curves of one telescope and the original light curve of a second telescope, finally the confidence bands can be determined. Here, the confidence limits are determined as the 1\%, 5\%, 95\% and 99\% quantiles of the 1000 resulting DCFs.

In the following plots, the black dots and error bars are the DCF and its error calculated after \citet{1988ApJ...333..646E}. The blue and green lines represent the confidence limits of 95\% and 5\% and of 99\% and 1\% respectively determined with DCFs of the 1000 simulated light curves of the first telescope and the original light curve of the second telescope. A value above the 99\% confidence limit is considered as a significant correlation, a significant anti-correlation is given for a value below the 1\% limit.

A binning of two days is chosen in this case. The reason for this is the unequal binning of the light curves which might lead to shifts in the correlations by one day when the time difference in the two light curves is larger than half a day. Time lags between $-50$ and +50 days are examined. The time lag $\Delta t$ is defined as the time difference of the second light curve to the first light curve (Instrument$_\text{1}$ vs. Instrument$_\text{2}$).\\

In the following subsections we report the results from our study on the correlation between the optical, X-ray and VHE $\gamma$-ray bands. The radio bands do not show significant variability and hence the radio fluxes cannot be correlated to the fluxes in the other bands.

\subsection{\textit{RXTE}/ASM and MAGIC}
\label{DCFRXTEMAGIC}

\begin{figure}[tb]
\centering
    \includegraphics[width=0.49\textwidth]{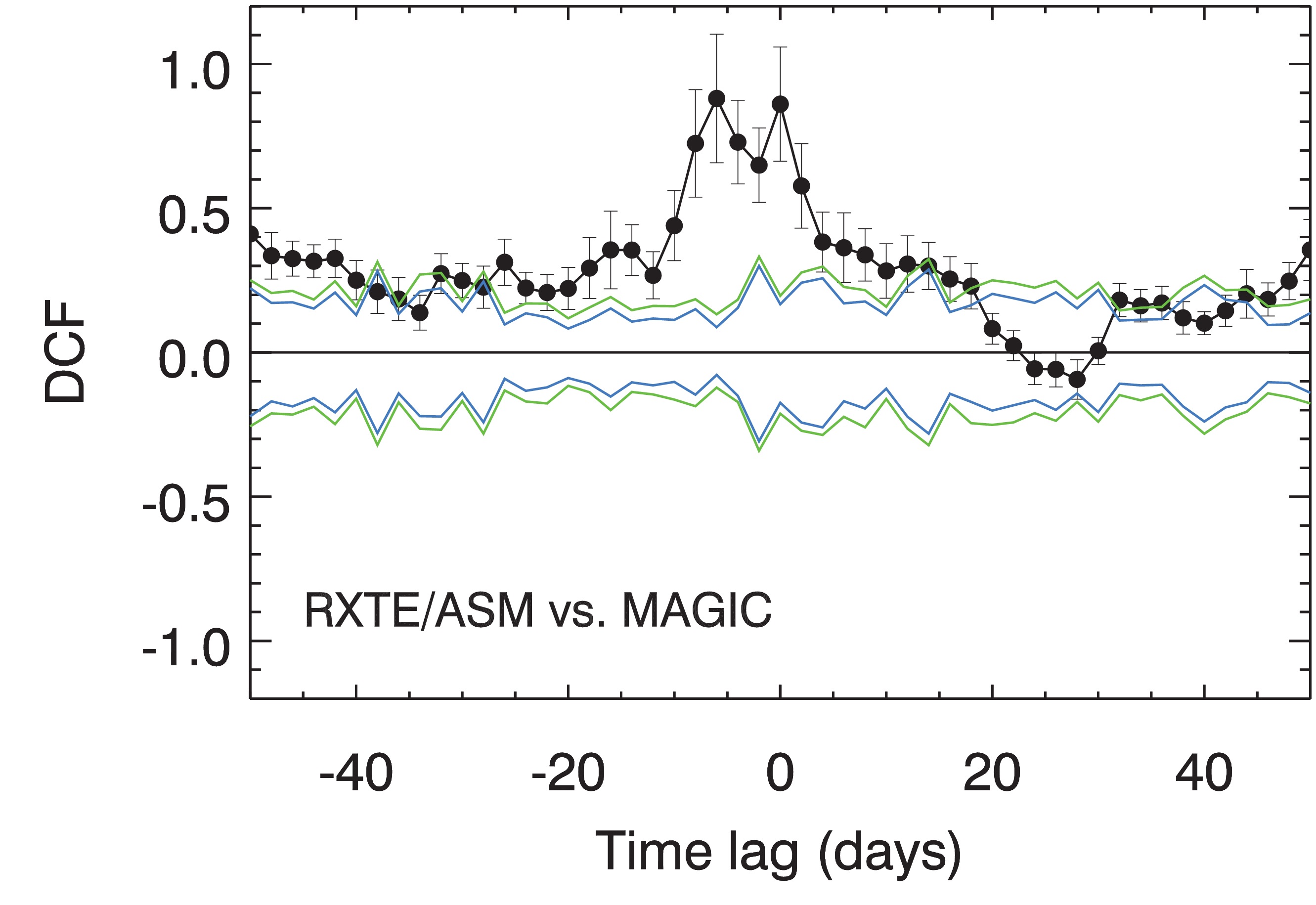}  
    \caption{Discrete Correlation Function for the light curves of \textit{RXTE}/ASM and MAGIC for the 2.3 year long period (Period~1, 2 and 3). Time lags from -50 to +50 days  in steps of 2 days are considered. The black dots represent the DCF values with the error bars calculated as in \citet{1988ApJ...333..646E}. The green (blue) lines represent the 99\% and 1\% (95\% and 5\%) confidence limits for random correlations resulting from the dedicated Monte Carlo analysis described in section \ref{sec:Correlations}.}
    \label{fig:dcf_rxteall_magicall}
\end{figure}

The \textit{RXTE}/ASM and MAGIC cross-correlations were examined at first for the whole time range from February 2007 to June 2009. This is reported in Figure~\ref{fig:dcf_rxteall_magicall}. 

There is positive and significant correlation for the entire range of time lags considered, that is from $-50$ to +50 days. The main cause of this positive correlation is the substantially larger flux level in Period~2, in comparison to that in Periods~1 and 3. If the light curves are shifted by a time lag smaller than the duration of these periods (e.g. 50 days), the pairing of VHE $\gamma$-ray fluxes and X-ray fluxes occurs always (for all time lags) within the observations from the same period, and hence one gets high VHE flux values related to high X-ray flux values, i.e. all from Period~2, and low VHE flux values matched with low X-ray flux values, i.e. all from Periods~1 and 3. And this effect naturally produces a positive correlation.\\

\begin{figure}[tb]
\centering
    \includegraphics[width=0.49\textwidth]{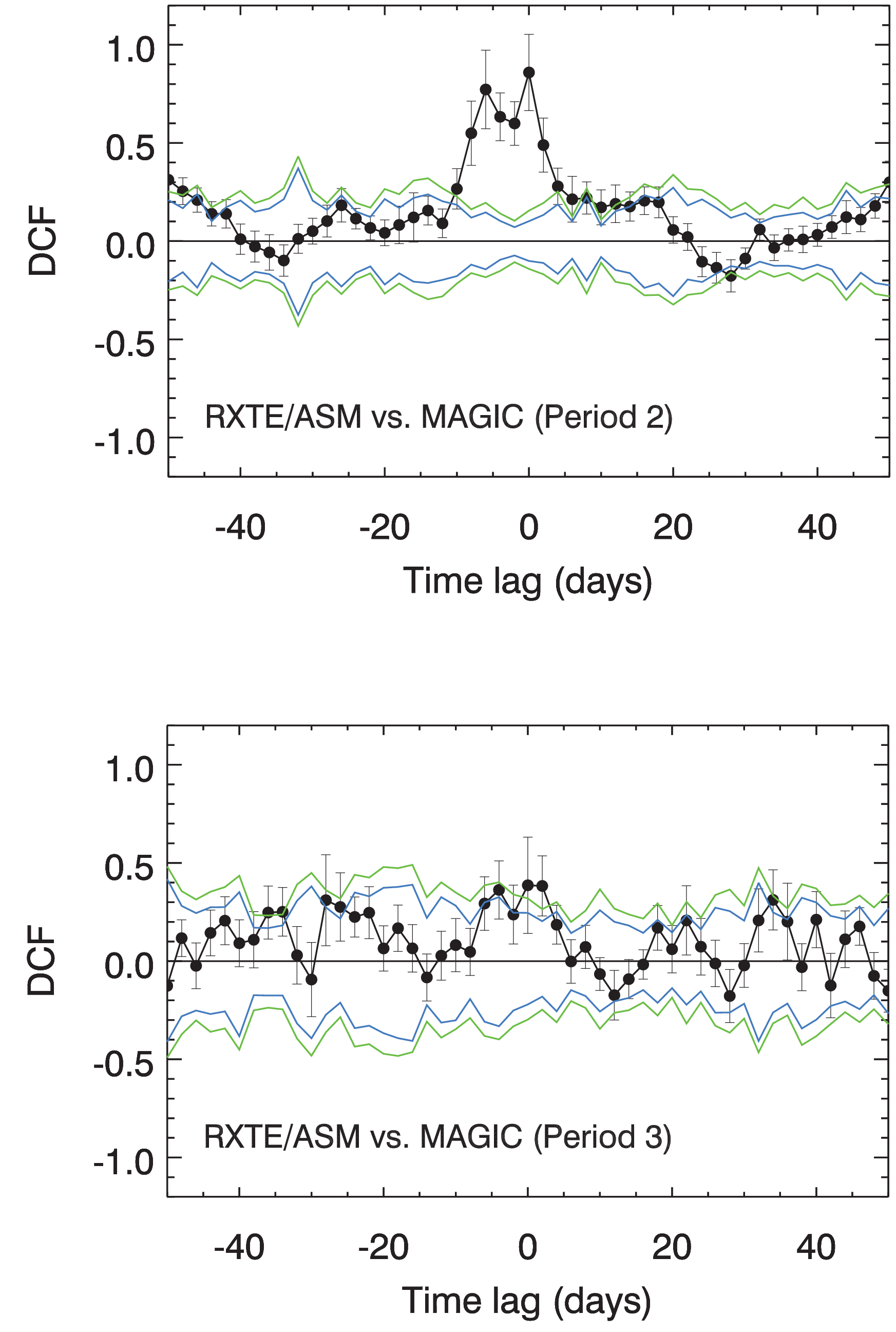}  
    \caption{Discrete Correlation Function for the light curves of \textit{RXTE}/ASM and MAGIC for Period~2 (top) and for Period~3 (bottom). The description of data points and contours are given in the caption of Fig. \ref{fig:dcf_rxteall_magicall}.}
    \label{fig:dcf_rxte_magic_P23}
\end{figure}

To remove the effect of the substantially different flux levels between the different periods, as well as to test the influence of the different states of activity and flux strength reported in the previous sections, the DCF is determined separately for Periods~2 and 3. The MAGIC light curve in the quiet Period~1 contains only six data points and is therefore not included in this study. The results are shown in Figure~\ref{fig:dcf_rxte_magic_P23}. We note that there is still an overall positive correlation for both Periods~2 and 3, however the DCF values are typically within the 95\% confidence contours. This positive (but not significant) correlation for all time lags is due to the fact that the two light curves considered here have the same overall trends: in Period~2 the VHE $\gamma$-ray and the X-ray light curves show an overall flux increase throughout the entire period, whereas in Period~3 they both show an overall decrease.\\

The quiet Period~3 shows a marginally significant correlation around a time lag of zero, while the active Period~2 shows a prominent correlation, with some structure around a time lag of zero. The DCF structure depicted in the top panel of Figure~\ref{fig:dcf_rxte_magic_P23} resembles that in Figure~\ref{fig:dcf_rxteall_magicall}, which indicates that the correlations in the high-activity Period~2 dominate the DCF values reported in Figure~\ref{fig:dcf_rxteall_magicall}, which relate to the full 2.3 years time interval. In both cases, one finds a peak at $\Delta t$= 0 and $\Delta t$= $-6$ days. The first peak is due to the direct correlation dominated by simultaneous prominent features in both light curves (i.e. flares on MJD~54556 and 54622). On the other hand, the DCF peak at $-6$ days is dominated by the remarkable 3-day long X-ray flaring activity around MJD~54630, which is the highest flux value in the \textit{RXTE}/ASM light curve. There is no counterpart in the VHE $\gamma$-ray light curve because MAGIC did not observe around that date, but this prominent X-ray flaring activity is matched with the large VHE flaring activity around MJD~54622 for time lags of around $-6$ days. The relatively broad structure of positive DCF values, extending from $-10$ days to +6 days, is dominated by the remarkable and asymmetric flaring activity in the X-ray light curve in a broad region around MJD~54556, which is coindicent with the relatively short VHE flare at the same location. 

\subsection{\textit{Swift}/BAT and MAGIC}
\label{subsec:DCFMAGICSwift}

\begin{figure}[tb]
\centering
    \includegraphics[width=0.49\textwidth]{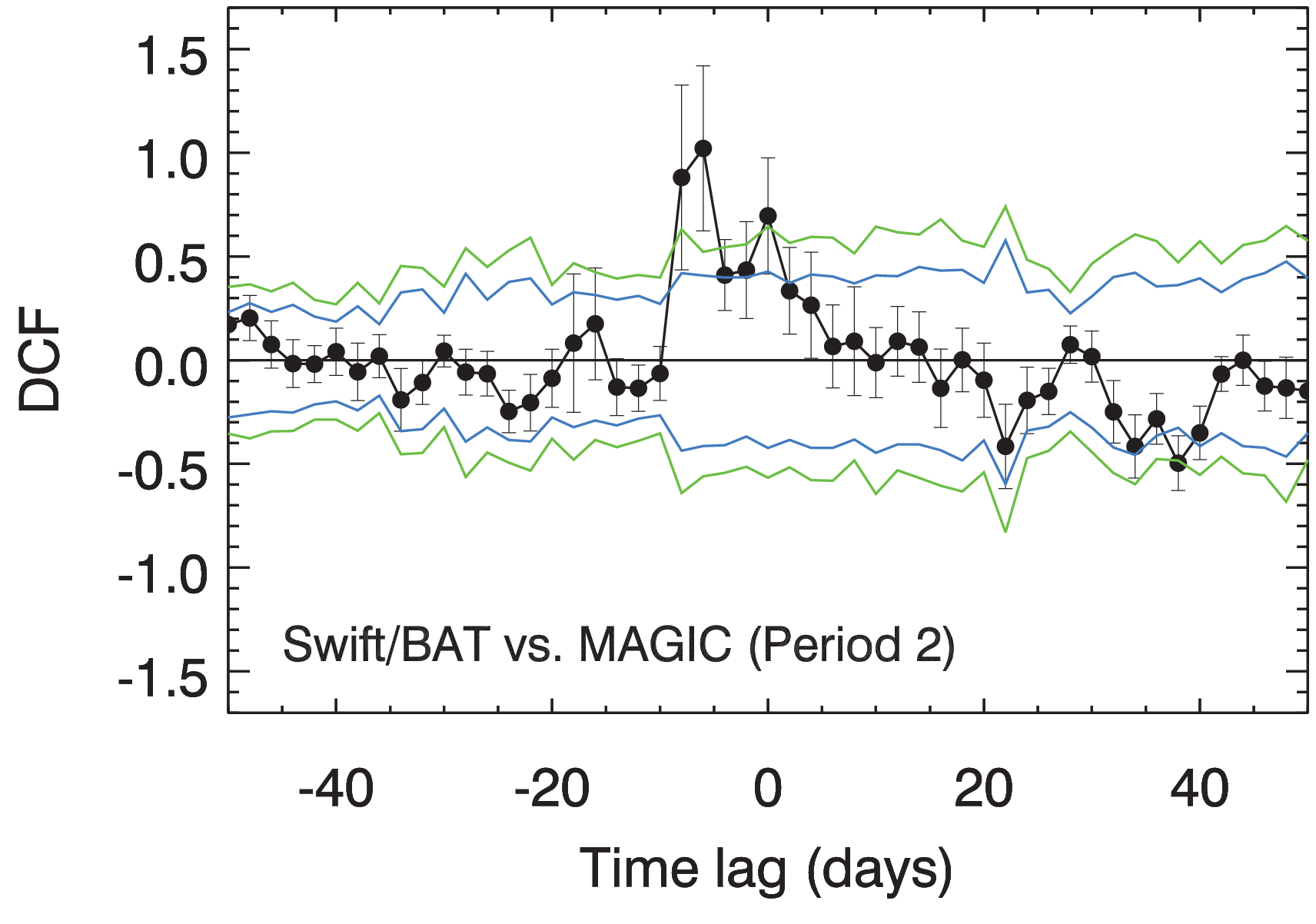}  
    \caption{Discrete Correlation Function for the light curves of \textit{Swift}/BAT and MAGIC for Period~2. The description of data points and contours are given in the caption of Fig. \ref{fig:dcf_rxteall_magicall}.}
    \label{fig:dcf_swift2008_magic2008}
\end{figure}

The sensitivity and temporal coverage of \textit{Swift}/BAT is somewhat lower than that of \textit{RXTE}/ASM, which reduces the accuracy with which one can study the correlation between the hard X-ray band above 15\,keV and the VHE $\gamma$-rays. For Period~3, we only could find a marginally significant correlation dominated by the somewhat higher X-ray and VHE flux values in the MJD range from 54858 to 54864. In Figure \ref{fig:dcf_swift2008_magic2008} the correlation results of the \textit{Swift}/BAT and the MAGIC light curves in the high-activity Period~2 are shown. When considering this period, we find DCF values above the 95\% confidence level for time lags between $-8$ days and +2 days, with two peaks above the 99\% confidence level for the time lags of 0, and also $-8$ and $-6$ days. The explanation of these two peaks is essentially the same that was given for the correlations between \textit{RXTE}/ASM and MAGIC reported in Section~\ref{DCFRXTEMAGIC}. The peak at $\Delta t$= 0 is dominated by several features appearing simultaneously in both light curves, the peak at $-6$ to $-8$ days is dominated by the large 3-day long X-ray activity around MJD~54630 (where we do not have MAGIC observations), and the broad and somewhat asymmetric structure in the DCF plot is dominated by the large and broad and asymmetric X-ray flaring around MJD~54556.

\subsection{GASP-WEBT and MAGIC}

\begin{figure}[tb]
\centering
    \includegraphics[width=0.49\textwidth]{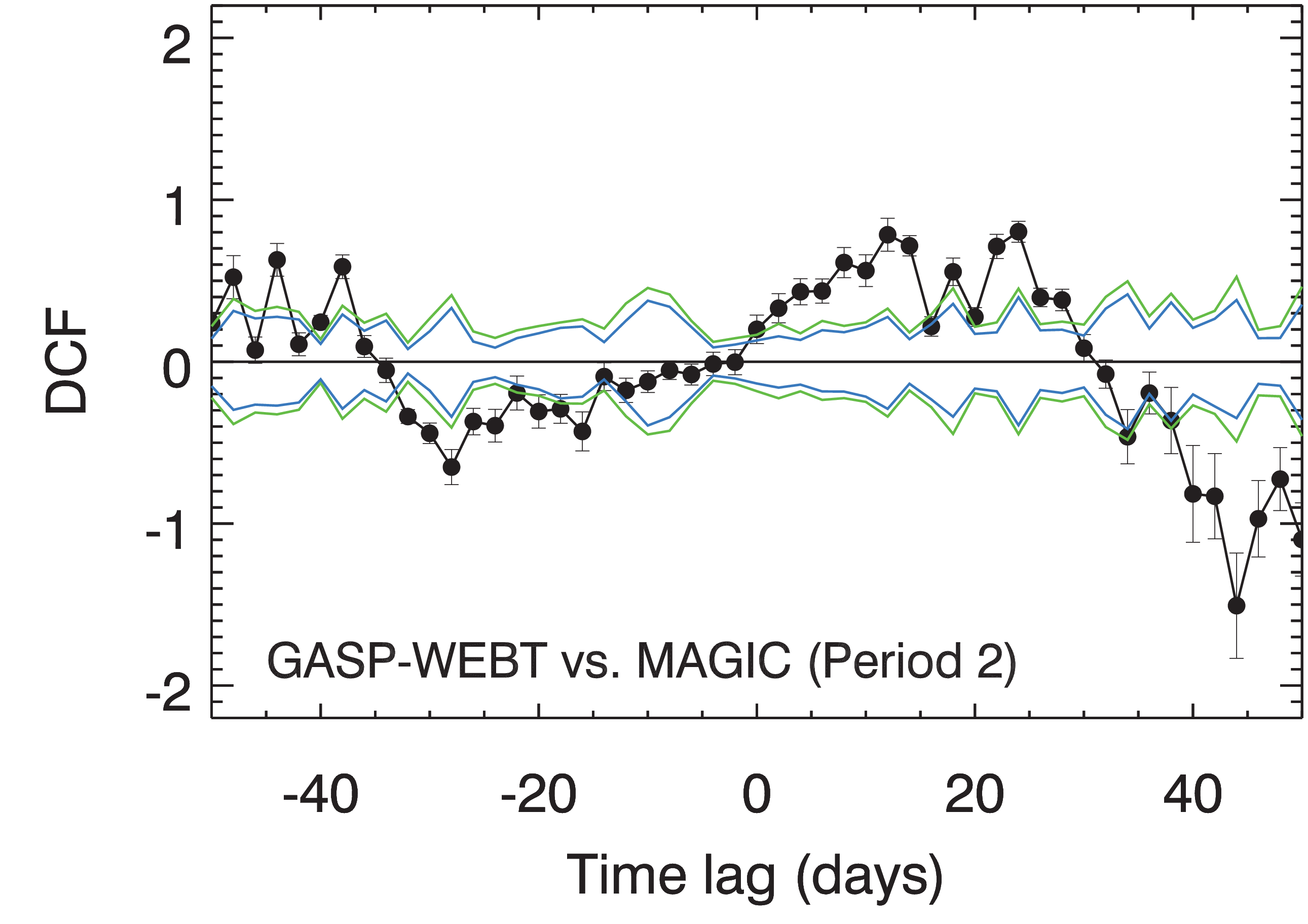}  
    \caption{Discrete Correlation Function for the light curves of GASP-WEBT and MAGIC for Period~2. The description of data points and contours are given in the caption of Fig. \ref{fig:dcf_rxteall_magicall}.}
    \label{fig:dcf_gaspwebt2008_magic2008}
\end{figure}

The correlation between the GASP-WEBT and MAGIC light curve for the high-activity Period~2 is shown in Figure~\ref{fig:dcf_gaspwebt2008_magic2008}. There is a positive correlation for time lags between 0 and +28 days, as well as around $-44$ days, and a negative correlation for time lags around $-28$ and around +44 days. This alternation of correlation and anti-correlation is caused by the fact that the variability in the optical and VHE emission is dominated by two to three prominent features. And hence the alternating presence of rises and drops in flux in both light curves creates these features in the DCF. For instance, when  shifting the optical light curve by e.g. +24 days or $-44$ days, minima and maxima in both light curves get aligned yielding a significant correlation, while when the optical light curve is shifted by $-28$ or +44 days, the minima in one light curve are aligned with maxima in the other light curve, hence yielding a significant anti-correlation. Although the reported correlations for some time lags are significant from the statistical point of view, they are based on the alignment or misalignment of only two to three prominent and relatively broad features, and these prominent features are not necessarily related to each other.

In the quiet Period~3, we find an overall anti-correlation during the entire range of time lags proved. This result is produced by the overall flux decrease in the VHE light curve and the overall flux increase in the optical light curve throughout the entire Period~3. The same result was reported and discussed in \citet{2015A&A...576A.126A}.

\subsection{GASP-WEBT and \textit{RXTE}/ASM}

\begin{figure}[tb]
\centering
    \includegraphics[width=0.49\textwidth]{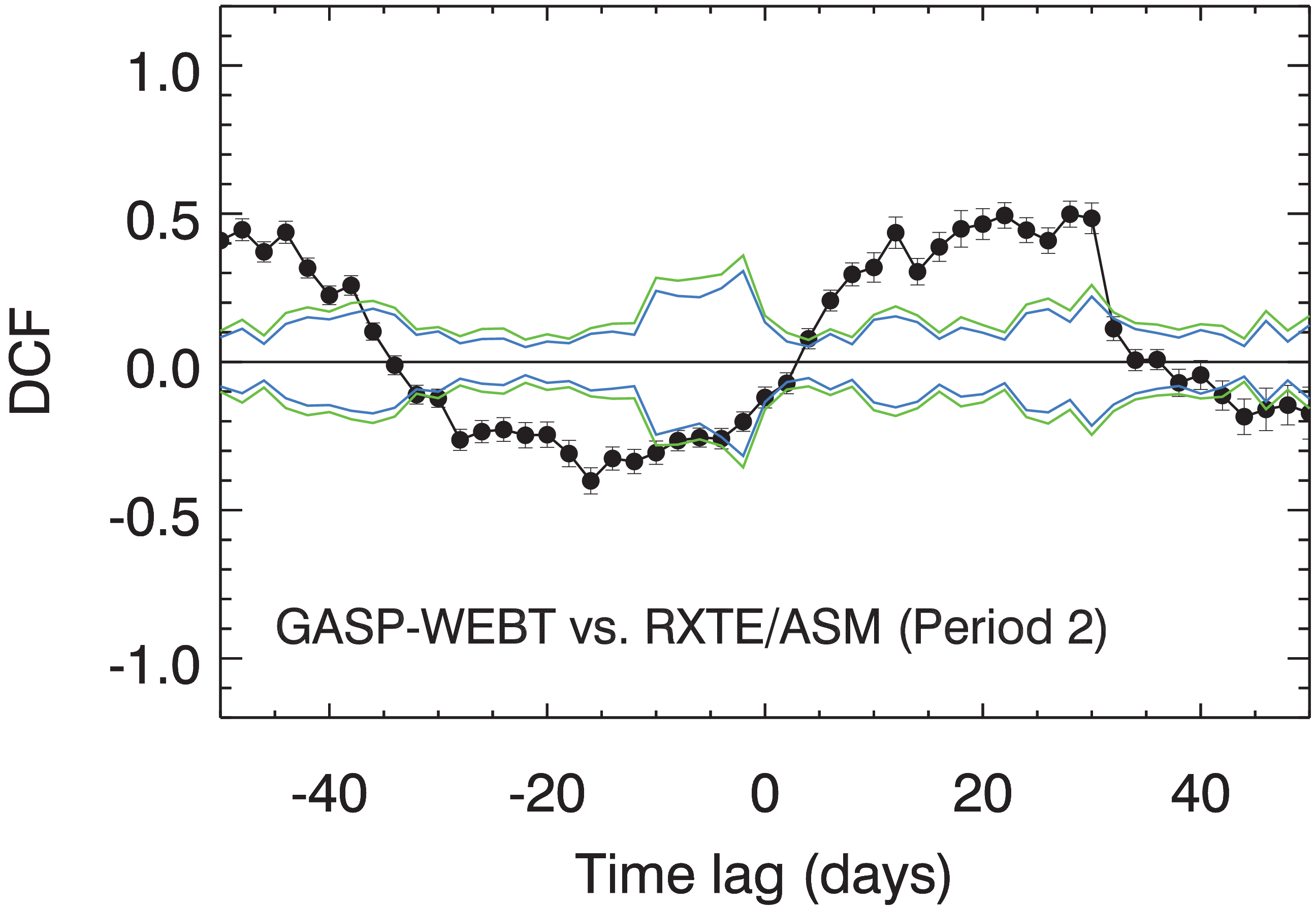}
    \caption{Discrete Correlation Function for the light curves of GASP-WEBT and \textit{RXTE}/ASM for Period~2. The description of data points and contours are given in the caption of Fig. \ref{fig:dcf_rxteall_magicall}.}
    \label{fig:dcf_gaspwebt2008_rxte2008}
\end{figure}

The DCF results of GASP-WEBT and \textit{RXTE}/ASM in Period~2 are shown in Figure~\ref{fig:dcf_gaspwebt2008_rxte2008}. A correlation is seen for positive time lags between +6 and +30 days, as well as for negative time lags between $-50$ and $-38$ days. Anti-correlations are seen between $-28$ and $-10$ days and between +44 and +50 days. These results are comparable to the results between GASP-WEBT and MAGIC. This again shows the alternation of rises and drops in flux produced by the fact that the variability in the optical and X-ray emission is dominated by only two to three prominent features. When shifting the optical light curve by the time lags, for which correlations are found, maxima in both light curves are aligned. When shifting the light curve by the time lags, for which anti-correlations are found, minima in the optical light curve are aligned with maxima in the X-ray light curves. Again, these correlations and anti-correlations might have been found by chance.

In Period~1 no correlations nor anti-correlations are seen for this pair of instruments. However, in Period~3 the GASP-WEBT light curve shows an overall anti-correlation with the \textit{RXTE}/ASM light curve, which occurs due to the overall slow decrease of the X-ray rate and the flux increase in the optical light curve. This result is comparable to the overall anti-correlation for the X-ray and TeV $\gamma$-ray light curves discussed in \citet{2015A&A...576A.126A}, which used partially the same data set.

\section{Summary of Results}
\label{sec:Summary}
\begin{enumerate}[i]

\item Between March 2007 and June 2009, MAGIC-I accumulated 96~hours of VHE $\gamma$-ray data of the blazar Mrk~421: the VHE flux varied around the typical flux baseline of about 0.5\,CU, with the highest flux of about 3.8\,CU occurring during the active state in 2008. 

\item For the first time the Bayesian Block algorithm was applied to the VHE $\gamma$-ray light curve from a Cherenkov telescope to identify different flux emission states, as well as to quantify the variability and to search for flaring activity.

\item The MAGIC $\gamma$-ray light curve was compared to light curves of other wavebands, including the hard and soft X-ray wavebands from \textit{Swift}BAT and \textit{RXTE}/ASM, the optical R-Band from GASP-WEBT, and two radio wavebands from Mets\"{a}hovi and OVRO.

\item The VHE and X-ray light curves resemble each other, showing a number of few-day long structures, while the optical and radio light curves show smaller flux variations and occurring on longer time scales.

\item  The fractional variability is low for radio and optical wavebands, and high for the X-ray and VHE
$\gamma$-ray bands during both low and high activity.

\item The discrete correlation function shows a direct relation between the two X-ray bands and the VHE $\gamma$-ray band, while no correlation was found between the optical and the X-ray and VHE bands.

\end{enumerate}

\section{Discussion and Conclusions}
\label{sec:Discussion}

A comprehensive variability and correlation study has been performed with 2.3 years of multi-band data from Mrk 421. The measured variability as a function of energy, with the highest variability in the X-ray and VHE bands, and the observed direct X-ray-to-VHE correlation, both occuring comparably during high- and low-activity, suggests that the processes that dominate the flux variability in Mrk~421 are similar for the different activity levels. The pattern characterized by a high variability in the X-ray and $\gamma$-ray emission, accompanied by a low variability in the optical and radio emission, occurs in both quiescent and excited states, qualifying this behaviour as typical of Mrk~421. The low variability and different time scales observed both in the radio and optical emission may be explained by different emission regions, or by cooler electrons in the jet at a later time. Additionally, the correlation between the X-ray and the VHE $\gamma$-ray emission extending over many months suggests that the broadband emission of Mrk~421 is predominantly produced by the same particles, e.g. via the Synchrotron-Self-Compton process. Alternatively, the X-rays and $\gamma$-rays could both result from the same radiation process (e.g. synchrotron radiation), but from two different electron populations varying together most times, but not necessarily always. This is the case in hadronic scenarios where the X-ray and $\gamma$-ray photons result from the synchrotron radiation of electrons in subsequent and therefore coupled cascade generations (\citet{1993A&A...269...67M}). The cascade generations are driven by the pair production in photon-photon scatterings involving low-energy photon fields, which can vary themselves, thereby modulating the variations of flux of the primary photo-mesons at the top of the cascades.

\begin{acknowledgements}
We would like to thank the Instituto de Astrof\'{\i}sica de Canarias for the excellent working conditions at the Observatorio del Roque de los Muchachos in La Palma. The financial support of the German BMBF and MPG, the Italian INFN and INAF, the Swiss National Fund SNF, the ERDF under the Spanish MINECO (FPA2012-39502), and the Japanese JSPS and MEXT is gratefully acknowledged. This work was also supported by the Centro de Excelencia Severo Ochoa SEV-2012-0234, CPAN CSD2007-00042, and MultiDark CSD2009-00064 projects of the Spanish Consolider-Ingenio 2010 programme, by grant 268740 of the Academy of Finland, by the Croatian Science Foundation (HrZZ) Project 09/176 and the University of Rijeka Project 13.12.1.3.02, by the DFG Collaborative Research Centers SFB823/C4 and SFB876/C3, and by the Polish MNiSzW grant 745/N-HESS-MAGIC/2010/0.\\

The public data archives of \textit{Swift}/BAT and \textit{RXTE}/ASM are acknowledged. \\

We thank the OVRO telescope for making its results available for the public. The OVRO 40\,m monitoring program is supported in part by NASA grants NNX08AW31G and NNX11A043G, and NFS grants AST-0808050 and AST-1109911.\\

We also thank the KVA and Mets\"ahovi telescopes for making their light curves available. M. Villata organized the optical-to-radio observations by GASP-WEBT as the president of the collaboration.\\

The Mets\"ahovi team acknowledges the support from the Academy of Finland to our observing projects (numbers 212656, 210338, 121148, and others). St. Petersburg University team acknowledges support from Russian RFBR grant 15-02-00949 and St. Petersburg University research grant 6.38.335.2015. The Abastumani Observatory team acknowledges financial support by the Shota Rustaveli National Science Foundation under contract FR/577/6-320/13.

\end{acknowledgements}

\bibliographystyle{aa} 
\bibliography{Mrk421_LongTerm_20072009}

\begin{thebibliography}{52}
\expandafter\ifx\csname natexlab\endcsname\relax\def\natexlab#1{#1}\fi

\bibitem[{{Abdo} {et~al.}(2011){Abdo}, {Ackermann}, {Ajello}, {Baldini},
  {Ballet}, {Barbiellini}, {Bastieri}, {Bechtol}, {Bellazzini}, {Berenji}, \&
  et~al.}]{2011ApJ...736..131A}
{Abdo}, A.~A., {Ackermann}, M., {Ajello}, M., {et~al.} 2011, \apj, 736, 131

\bibitem[{{Acciari} {et~al.}(2011){Acciari}, {Aliu}, {Arlen}, {Aune},
  {Beilicke}, {Benbow}, {Boltuch}, {Bradbury}, {Buckley}, {Bugaev}, {Byrum},
  {Cannon}, {Cesarini}, {Ciupik}, {Cui}, {Dickherber}, {Duke}, {Falcone},
  {Finley}, {Finnegan}, {Fortson}, {Furniss}, {Galante}, {Gall}, {Gillanders},
  {Godambe}, {Grube}, {Guenette}, {Gyuk}, {Hanna}, {Holder}, {Hui}, {Humensky},
  {Imran}, {Kaaret}, {Karlsson}, {Kertzman}, {Kieda}, {Konopelko},
  {Krawczynski}, {Krennrich}, {Lang}, {Maier}, {McArthur}, {McCutcheon},
  {Moriarty}, {Ong}, {Otte}, {Ouellette}, {Pandel}, {Perkins}, {Pichel},
  {Pohl}, {Quinn}, {Ragan}, {Reyes}, {Reynolds}, {Roache}, {Rose}, {Rovero},
  {Schroedter}, {Sembroski}, {Senturk}, {Steele}, {Swordy}, {Theiling},
  {Thibadeau}, {Varlotta}, {Vassiliev}, {Vincent}, {Wagner}, {Wakely}, {Ward},
  {Weekes}, {Weinstein}, {Weisgarber}, {Williams}, {Wissel}, {Wood}, {Zitzer},
  {Garson}, {Lee}, {Sadun}, {Carini}, {Barnaby}, {Cook}, {Maune}, {Pease},
  {Smith}, {Walters}, {Berdyugin}, {Lindfors}, {Nilsson}, {Pasanen}, {Sainio},
  {Sillanpaa}, {Takalo}, {Villforth}, {Montaruli}, {Baker}, {Lahteenmaki},
  {Tornikoski}, {Hovatta}, {Nieppola}, {Aller}, \&
  {Aller}}]{2011ApJ...738...25A}
{Acciari}, V.~A., {Aliu}, E., {Arlen}, T., {et~al.} 2011, \apj, 738, 25

\bibitem[{{Acciari} {et~al.}(2009){Acciari}, {Aliu}, {Aune}, {Beilicke},
  {Benbow}, {B{\"o}ttcher}, {Bradbury}, {Buckley}, {Bugaev}, {Butt}, \&
  et~al.}]{2009ApJ...703..169A}
{Acciari}, V.~A., {Aliu}, E., {Aune}, T., {et~al.} 2009, \apj, 703, 169

\bibitem[{{Acciari} {et~al.}(2014){Acciari}, {Arlen}, {Aune}, {Benbow}, {Bird},
  {Bouvier}, {Bradbury}, {Buckley}, {Bugaev}, {de la Calle Perez},
  {Carter-Lewis}, {Cesarini}, {Ciupik}, {Collins-Hughes}, {Connolly}, {Cui},
  {Duke}, {Dumm}, {Falcone}, {Federici}, {Fegan}, {Fegan}, {Finley},
  {Finnegan}, {Fortson}, {Gaidos}, {Galante}, {Gall}, {Gibbs}, {Gillanders},
  {Griffin}, {Grube}, {Gyuk}, {Hanna}, {Horan}, {Humensky}, {Kaaret},
  {Kertzman}, {Khassen}, {Kieda}, {Krawczynski}, {Krennrich}, {Lang},
  {McEnery}, {Madhavan}, {Moriarty}, {Nelson}, {O'Faol{\'a}in de Bhr{\'o}ithe},
  {Ong}, {Orr}, {Otte}, {Perkins}, {Petry}, {Pichel}, {Pohl}, {Quinn}, {Ragan},
  {Reynolds}, {Roache}, {Rovero}, {Schroedter}, {Sembroski}, {Smith},
  {Telezhinsky}, {Theiling}, {Toner}, {Tyler}, {Varlotta}, {Vivier}, {Wakely},
  {Ward}, {Weekes}, {Weinstein}, {Welsing}, {Williams}, \&
  {Wissel}}]{2014APh....54....1A}
{Acciari}, V.~A., {Arlen}, T., {Aune}, T., {et~al.} 2014, Astroparticle
  Physics, 54, 1

\bibitem[{{Albert} {et~al.}(2007){Albert}, {Aliu}, {Anderhub}, {Antoranz},
  {Armada}, {Asensio}, {Baixeras}, {Barrio}, {Bartko}, {Bastieri}, {Becker},
  {Bednarek}, {Berger}, {Bigongiari}, {Biland}, {Bock}, {Bordas},
  {Bosch-Ramon}, {Bretz}, {Britvitch}, {Camara}, {Carmona}, {Chilingarian},
  {Ciprini}, {Coarasa}, {Commichau}, {Contreras}, {Cortina}, {Curtef},
  {Danielyan}, {Dazzi}, {De Angelis}, {de los Reyes}, {De Lotto},
  {Domingo-Santamar{\'{\i}}a}, {Dorner}, {Doro}, {Errando}, {Fagiolini},
  {Ferenc}, {Fern{\'a}ndez}, {Firpo}, {Flix}, {Fonseca}, {Font}, {Fuchs},
  {Galante}, {Garczarczyk}, {Gaug}, {Giller}, {Goebel}, {Hakobyan},
  {Hayashida}, {Hengstebeck}, {H{\"o}hne}, {Hose}, {Hsu}, {Jacon}, {Jogler},
  {Kalekin}, {Kosyra}, {Kranich}, {Kritzer}, {Laatiaoui}, {Laille}, {Liebing},
  {Lindfors}, {Lombardi}, {Longo}, {L{\'o}pez}, {L{\'o}pez}, {Lorenz},
  {Majumdar}, {Maneva}, {Mannheim}, {Mansutti}, {Mariotti}, {Mart{\'{\i}}nez},
  {Mazin}, {Merck}, {Meucci}, {Meyer}, {Miranda}, {Mirzoyan}, {Mizobuchi},
  {Moralejo}, {Nilsson}, {Ninkovic}, {O{\~n}a-Wilhelmi}, {Ordu{\~n}a}, {Otte},
  {Oya}, {Paneque}, {Paoletti}, {Paredes}, {Pasanen}, {Pascoli}, {Pauss},
  {Pegna}, {Persic}, {Peruzzo}, {Piccioli}, {Poller}, {Prandini}, {Raymers},
  {Rhode}, {Rib{\'o}}, {Rico}, {Rissi}, {Robert}, {R{\"u}gamer}, {Saggion},
  {S{\'a}nchez}, {Sartori}, {Scalzotto}, {Scapin}, {Schmitt}, {Schweizer},
  {Shayduk}, {Shinozaki}, {Shore}, {Sidro}, {Sillanp{\"a}{\"a}}, {Sobczynska},
  {Stamerra}, {Stark}, {Takalo}, {Temnikov}, {Tescaro}, {Teshima}, {Tonello},
  {Torres}, {Torres}, {Turini}, {Vankov}, {Vitale}, {Wagner}, {Wibig},
  {Wittek}, {Zanin}, \& {Zapatero}}]{2007ApJ...663..125A}
{Albert}, J., {Aliu}, E., {Anderhub}, H., {et~al.} 2007, \apj, 663, 125

\bibitem[{{Albert} {et~al.}(2008){Albert}, {Aliu}, {Anderhub}, {Antoranz},
  {Armada}, {Baixeras}, {Barrio}, {Bartko}, {Bastieri}, {Becker}, {Bednarek},
  {Berger}, {Bigongiari}, {Biland}, {Bock}, {Bordas}, {Bosch-Ramon}, {Bretz},
  {Britvitch}, {Camara}, {Carmona}, {Chilingarian}, {Coarasa}, {Commichau},
  {Contreras}, {Cortina}, {Costado}, {Curtef}, {Danielyan}, {Dazzi}, {De
  Angelis}, {Delgado}, {de los Reyes}, {De Lotto}, {Domingo-Santamar{\'{\i}}a},
  {Dorner}, {Doro}, {Errando}, {Fagiolini}, {Ferenc}, {Fern{\'a}ndez}, {Firpo},
  {Flix}, {Fonseca}, {Font}, {Fuchs}, {Galante}, {Garc{\'{\i}}a-L{\'o}pez},
  {Garczarczyk}, {Gaug}, {Giller}, {Goebel}, {Hakobyan}, {Hayashida},
  {Hengstebeck}, {Herrero}, {H{\"o}hne}, {Hose}, {Hsu}, {Jacon}, {Jogler},
  {Kosyra}, {Kranich}, {Kritzer}, {Laille}, {Lindfors}, {Lombardi}, {Longo},
  {L{\'o}pez}, {L{\'o}pez}, {Lorenz}, {Majumdar}, {Maneva}, {Mannheim},
  {Mansutti}, {Mariotti}, {Mart{\'{\i}}nez}, {Mazin}, {Merck}, {Meucci},
  {Meyer}, {Miranda}, {Mirzoyan}, {Mizobuchi}, {Moralejo}, {Nieto}, {Nilsson},
  {Ninkovic}, {O{\~n}a-Wilhelmi}, {Otte}, {Oya}, {Paneque}, {Panniello},
  {Paoletti}, {Paredes}, {Pasanen}, {Pascoli}, {Pauss}, {Pegna}, {Persic},
  {Peruzzo}, {Piccioli}, {Poller}, {Prandini}, {Puchades}, {Raymers}, {Rhode},
  {Rib{\'o}}, {Rico}, {Rissi}, {Robert}, {R{\"u}gamer}, {Saggion},
  {S{\'a}nchez}, {Sartori}, {Scalzotto}, {Scapin}, {Schmitt}, {Schweizer},
  {Shayduk}, {Shinozaki}, {Shore}, {Sidro}, {Sillanp{\"a}{\"a}}, {Sobczynska},
  {Stamerra}, {Stark}, {Takalo}, {Temnikov}, {Tescaro}, {Teshima}, {Tonello},
  {Torres}, {Turini}, {Vankov}, {Vitale}, {Wagner}, {Wibig}, {Wittek},
  {Zandanel}, {Zanin}, \& {Zapatero}}]{2008ApJ...674.1037A}
{Albert}, J., {Aliu}, E., {Anderhub}, H., {et~al.} 2008, \apj, 674, 1037

\bibitem[{{Aleksi{\'c}} {et~al.}(2012){Aleksi{\'c}}, {Alvarez}, {Antonelli},
  {Antoranz}, {Asensio}, {Backes}, {Barrio}, {Bastieri}, {Becerra
  Gonz{\'a}lez}, {Bednarek}, {Berdyugin}, {Berger}, {Bernardini}, {Biland},
  {Blanch}, {Bock}, {Boller}, {Bonnoli}, {Borla Tridon}, {Braun}, {Bretz},
  {Ca{\~n}ellas}, {Carmona}, {Carosi}, {Colin}, {Colombo}, {Contreras},
  {Cortina}, {Cossio}, {Covino}, {Dazzi}, {De Angelis}, {De Caneva}, {De Cea
  del Pozo}, {De Lotto}, {Delgado Mendez}, {Diago Ortega}, {Doert},
  {Dom{\'{\i}}nguez}, {Dominis Prester}, {Dorner}, {Doro}, {Elsaesser},
  {Ferenc}, {Fonseca}, {Font}, {Fruck}, {Garc{\'{\i}}a L{\'o}pez},
  {Garczarczyk}, {Garrido}, {Giavitto}, {Godinovi{\'c}}, {Hadasch},
  {H{\"a}fner}, {Herrero}, {Hildebrand}, {H{\"o}hne-M{\"o}nch}, {Hose},
  {Hrupec}, {Huber}, {Jogler}, {Kellermann}, {Klepser}, {Kr{\"a}henb{\"u}hl},
  {Krause}, {La Barbera}, {Lelas}, {Leonardo}, {Lindfors}, {Lombardi},
  {L{\'o}pez}, {L{\'o}pez}, {Lorenz}, {Makariev}, {Maneva}, {Mankuzhiyil},
  {Mannheim}, {Maraschi}, {Mariotti}, {Mart{\'{\i}}nez}, {Mazin}, {Meucci},
  {Miranda}, {Mirzoyan}, {Miyamoto}, {Mold{\'o}n}, {Moralejo}, {Munar-Adrover},
  {Nieto}, {Nilsson}, {Orito}, {Oya}, {Paneque}, {Paoletti}, {Pardo},
  {Paredes}, {Partini}, {Pasanen}, {Pauss}, {Perez-Torres}, {Persic},
  {Peruzzo}, {Pilia}, {Pochon}, {Prada}, {Prada Moroni}, {Prandini}, {Puljak},
  {Reichardt}, {Reinthal}, {Rhode}, {Rib{\'o}}, {Rico}, {R{\"u}gamer},
  {Saggion}, {Saito}, {Saito}, {Salvati}, {Satalecka}, {Scalzotto}, {Scapin},
  {Schultz}, {Schweizer}, {Shayduk}, {Shore}, {Sillanp{\"a}{\"a}}, {Sitarek},
  {Sobczynska}, {Spanier}, {Spiro}, {Stamerra}, {Steinke}, {Storz}, {Strah},
  {Suri{\'c}}, {Takalo}, {Takami}, {Tavecchio}, {Temnikov}, {Terzi{\'c}},
  {Tescaro}, {Teshima}, {Tibolla}, {Torres}, {Treves}, {Uellenbeck}, {Vankov},
  {Vogler}, {Wagner}, {Weitzel}, {Zabalza}, {Zandanel}, \&
  {Zanin}}]{2012A&A...542A.100A}
{Aleksi{\'c}}, J., {Alvarez}, E.~A., {Antonelli}, L.~A., {et~al.} 2012, \aap,
  542, A100

\bibitem[{{Aleksi{\'c}} {et~al.}(2016{\natexlab{a}}){Aleksi{\'c}}, {Ansoldi},
  {Antonelli}, {Antoranz}, {Babic}, {Bangale}, {Barcel{\'o}}, {Barrio},
  {Becerra Gonz{\'a}lez}, {Bednarek}, {Bernardini}, {Biasuzzi}, {Biland},
  {Bitossi}, {Blanch}, {Bonnefoy}, {Bonnoli}, {Borracci}, {Bretz}, {Carmona},
  {Carosi}, {Cecchi}, {Colin}, {Colombo}, {Contreras}, {Corti}, {Cortina},
  {Covino}, {Da Vela}, {Dazzi}, {DeAngelis}, {De Caneva}, {De Lotto}, {de
  O{\~n}a Wilhelmi}, {Delgado Mendez}, {Dettlaff}, {Dominis Prester}, {Dorner},
  {Doro}, {Einecke}, {Eisenacher}, {Elsaesser}, {Fidalgo}, {Fink}, {Fonseca},
  {Font}, {Frantzen}, {Fruck}, {Galindo}, {Garc{\'{\i}}a L{\'o}pez},
  {Garczarczyk}, {Garrido Terrats}, {Gaug}, {Giavitto}, {Godinovi{\'c}},
  {Gonz{\'a}lez Mu{\~n}oz}, {Gozzini}, {Haberer}, {Hadasch}, {Hanabata},
  {Hayashida}, {Herrera}, {Hildebrand}, {Hose}, {Hrupec}, {Idec}, {Illa},
  {Kadenius}, {Kellermann}, {Knoetig}, {Kodani}, {Konno}, {Krause}, {Kubo},
  {Kushida}, {La Barbera}, {Lelas}, {Lemus}, {Lewandowska}, {Lindfors},
  {Lombardi}, {Longo}, {L{\'o}pez}, {L{\'o}pez-Coto}, {L{\'o}pez-Oramas},
  {Lorca}, {Lorenz}, {Lozano}, {Makariev}, {Mallot}, {Maneva}, {Mankuzhiyil},
  {Mannheim}, {Maraschi}, {Marcote}, {Mariotti}, {Mart{\'{\i}}nez}, {Mazin},
  {Menzel}, {Miranda}, {Mirzoyan}, {Moralejo}, {Munar-Adrover}, {Nakajima},
  {Negrello}, {Neustroev}, {Niedzwiecki}, {Nilsson}, {Nishijima}, {Noda},
  {Orito}, {Overkemping}, {Paiano}, {Palatiello}, {Paneque}, {Paoletti},
  {Paredes}, {Paredes-Fortuny}, {Persic}, {Poutanen}, {Prada Moroni},
  {Prandini}, {Puljak}, {Reinthal}, {Rhode}, {Rib{\'o}}, {Rico}, {Rodriguez
  Garcia}, {R{\"u}gamer}, {Saito}, {Saito}, {Satalecka}, {Scalzotto}, {Scapin},
  {Schultz}, {Schlammer}, {Schmidl}, {Schweizer}, {Sillanp{\"a}{\"a}},
  {Sitarek}, {Snidaric}, {Sobczynska}, {Spanier}, {Stamerra}, {Steinbring},
  {Storz}, {Strzys}, {Takalo}, {Takami}, {Tavecchio}, {Tejedor}, {Temnikov},
  {Terzi{\'c}}, {Tescaro}, {Teshima}, {Thaele}, {Tibolla}, {Torres}, {Toyama},
  {Treves}, {Vogler}, {Wetteskind}, {Will}, \& {Zanin}}]{2016APh....72...61A}
{Aleksi{\'c}}, J., {Ansoldi}, S., {Antonelli}, L.~A., {et~al.}
  2016{\natexlab{a}}, Astroparticle Physics, 72, 61

\bibitem[{{Aleksi{\'c}} {et~al.}(2016{\natexlab{b}}){Aleksi{\'c}}, {Ansoldi},
  {Antonelli}, {Antoranz}, {Babic}, {Bangale}, {Barcel{\'o}}, {Barrio},
  {Becerra Gonz{\'a}lez}, {Bednarek}, {Bernardini}, {Biasuzzi}, {Biland},
  {Bitossi}, {Blanch}, {Bonnefoy}, {Bonnoli}, {Borracci}, {Bretz}, {Carmona},
  {Carosi}, {Cecchi}, {Colin}, {Colombo}, {Contreras}, {Corti}, {Cortina},
  {Covino}, {Da Vela}, {Dazzi}, {De Angelis}, {De Caneva}, {De Lotto}, {de
  O{\~n}a Wilhelmi}, {Delgado Mendez}, {Dettlaff}, {Dominis Prester}, {Dorner},
  {Doro}, {Einecke}, {Eisenacher}, {Elsaesser}, {Fidalgo}, {Fink}, {Fonseca},
  {Font}, {Frantzen}, {Fruck}, {Galindo}, {Garc{\'{\i}}a L{\'o}pez},
  {Garczarczyk}, {Garrido Terrats}, {Gaug}, {Giavitto}, {Godinovi{\'c}},
  {Gonz{\'a}lez Mu{\~n}oz}, {Gozzini}, {Haberer}, {Hadasch}, {Hanabata},
  {Hayashida}, {Herrera}, {Hildebrand}, {Hose}, {Hrupec}, {Idec}, {Illa},
  {Kadenius}, {Kellermann}, {Knoetig}, {Kodani}, {Konno}, {Krause}, {Kubo},
  {Kushida}, {La Barbera}, {Lelas}, {Lemus}, {Lewandowska}, {Lindfors},
  {Lombardi}, {Longo}, {L{\'o}pez}, {L{\'o}pez-Coto}, {L{\'o}pez-Oramas},
  {Lorca}, {Lorenz}, {Lozano}, {Makariev}, {Mallot}, {Maneva}, {Mankuzhiyil},
  {Mannheim}, {Maraschi}, {Marcote}, {Mariotti}, {Mart{\'{\i}}nez}, {Mazin},
  {Menzel}, {Miranda}, {Mirzoyan}, {Moralejo}, {Munar-Adrover}, {Nakajima},
  {Negrello}, {Neustroev}, {Niedzwiecki}, {Nilsson}, {Nishijima}, {Noda},
  {Orito}, {Overkemping}, {Paiano}, {Palatiello}, {Paneque}, {Paoletti},
  {Paredes}, {Paredes-Fortuny}, {Persic}, {Poutanen}, {Prada Moroni},
  {Prandini}, {Puljak}, {Reinthal}, {Rhode}, {Rib{\'o}}, {Rico}, {Rodriguez
  Garcia}, {R{\"u}gamer}, {Saito}, {Saito}, {Satalecka}, {Scalzotto}, {Scapin},
  {Schultz}, {Schlammer}, {Schmidl}, {Schweizer}, {Shore}, {Sillanp{\"a}{\"a}},
  {Sitarek}, {Snidaric}, {Sobczynska}, {Spanier}, {Stamerra}, {Steinbring},
  {Storz}, {Strzys}, {Takalo}, {Takami}, {Tavecchio}, {Tejedor}, {Temnikov},
  {Terzi{\'c}}, {Tescaro}, {Teshima}, {Thaele}, {Tibolla}, {Torres}, {Toyama},
  {Treves}, {Vogler}, {Wetteskind}, {Will}, \& {Zanin}}]{2016APh....72...76A}
{Aleksi{\'c}}, J., {Ansoldi}, S., {Antonelli}, L.~A., {et~al.}
  2016{\natexlab{b}}, Astroparticle Physics, 72, 76

\bibitem[{{Aleksi{\'c}} {et~al.}(2015{\natexlab{a}}){Aleksi{\'c}}, {Ansoldi},
  {Antonelli}, {Antoranz}, {Babic}, {Bangale}, {Barres de Almeida}, {Barrio},
  {Becerra Gonz{\'a}lez}, {Bednarek}, {Berger}, {Bernardini}, {Biland},
  {Blanch}, {Bock}, {Bonnefoy}, {Bonnoli}, {Borracci}, {Bretz}, {Carmona},
  {Carosi}, {Carreto Fidalgo}, {Colin}, {Colombo}, {Contreras}, {Cortina},
  {Covino}, {Da Vela}, {Dazzi}, {De Angelis}, {De Caneva}, {De Lotto}, {Delgado
  Mendez}, {Doert}, {Dom{\'{\i}}nguez}, {Dominis Prester}, {Dorner}, {Doro},
  {Einecke}, {Eisenacher}, {Elsaesser}, {Farina}, {Ferenc}, {Fonseca}, {Font},
  {Frantzen}, {Fruck}, {Garc{\'{\i}}a L{\'o}pez}, {Garczarczyk}, {Garrido
  Terrats}, {Gaug}, {Giavitto}, {Godinovi{\'c}}, {Gonz{\'a}lez Mu{\~n}oz},
  {Gozzini}, {Hadamek}, {Hadasch}, {Herrero}, {Hildebrand}, {Hose}, {Hrupec},
  {Idec}, {Kadenius}, {Kellermann}, {Knoetig}, {Krause}, {Kushida}, {La
  Barbera}, {Lelas}, {Lewandowska}, {Lindfors}, {Longo}, {Lombardi},
  {L{\'o}pez}, {L{\'o}pez-Coto}, {L{\'o}pez-Oramas}, {Lorenz}, {Lozano},
  {Makariev}, {Mallot}, {Maneva}, {Mankuzhiyil}, {Mannheim}, {Maraschi},
  {Marcote}, {Mariotti}, {Mart{\'{\i}}nez}, {Mazin}, {Menzel}, {Meucci},
  {Miranda}, {Mirzoyan}, {Moralejo}, {Munar-Adrover}, {Nakajima},
  {Niedzwiecki}, {Nilsson}, {Nowak}, {Orito}, {Overkemping}, {Paiano},
  {Palatiello}, {Paneque}, {Paoletti}, {Paredes}, {Paredes-Fortuny}, {Partini},
  {Persic}, {Prada}, {Prada Moroni}, {Prandini}, {Preziuso}, {Puljak},
  {Reinthal}, {Rhode}, {Rib{\'o}}, {Rico}, {RodriguezGarcia}, {R{\"u}gamer},
  {Saggion}, {Saito}, {Salvati}, {Satalecka}, {Scalzotto}, {Scapin}, {Schultz},
  {Schweizer}, {Shore}, {Sillanp{\"a}{\"a}}, {Sitarek}, {Snidaric},
  {Sobczynska}, {Spanier}, {Stamatescu}, {Stamerra}, {Steinbring}, {Storz},
  {Sun}, {Suri{\'c}}, {Takalo}, {Tavecchio}, {Temnikov}, {Terzi{\'c}},
  {Tescaro}, {Teshima}, {Thaele}, {Tibolla}, {Torres}, {Toyama}, {Treves},
  {Uellenbeck}, {Vogler}, {Wagner}, {Zandanel}, \&
  {Zanin}}]{2015A&A...576A.126A}
{Aleksi{\'c}}, J., {Ansoldi}, S., {Antonelli}, L.~A., {et~al.}
  2015{\natexlab{a}}, \aap, 576, A126

\bibitem[{{Aleksi{\'c}} {et~al.}(2015{\natexlab{b}}){Aleksi{\'c}}, {Ansoldi},
  {Antonelli}, {Antoranz}, {Babic}, {Bangale}, {Barres de Almeida}, {Barrio},
  {Becerra Gonz{\'a}lez}, {Bednarek}, {Bernardini}, {Biasuzzi}, {Biland},
  {Blanch}, {Boller}, {Bonnefoy}, {Bonnoli}, {Borracci}, {Bretz}, {Carmona},
  {Carosi}, {Colin}, {Colombo}, {Contreras}, {Cortina}, {Covino}, {Da Vela},
  {Dazzi}, {De Angelis}, {De Caneva}, {De Lotto}, {de O{\~n}a Wilhelmi},
  {Delgado Mendez}, {Dominis Prester}, {Dorner}, {Doro}, {Einecke},
  {Eisenacher}, {Elsaesser}, {Fonseca}, {Font}, {Frantzen}, {Fruck}, {Galindo},
  {Garc{\'{\i}}a L{\'o}pez}, {Garczarczyk}, {Garrido Terrats}, {Gaug},
  {Godinovi{\'c}}, {Gonz{\'a}lez Mu{\~n}oz}, {Gozzini}, {Hadasch}, {Hanabata},
  {Hayashida}, {Herrera}, {Hildebrand}, {Hose}, {Hrupec}, {Hughes}, {Idec},
  {Kadenius}, {Kellermann}, {Knoetig}, {Kodani}, {Konno}, {Krause}, {Kubo},
  {Kushida}, {La Barbera}, {Lelas}, {Lewandowska}, {Lindfors}, {Lombardi},
  {L{\'o}pez}, {L{\'o}pez-Coto}, {L{\'o}pez-Oramas}, {Lorenz}, {Lozano},
  {Makariev}, {Mallot}, {Maneva}, {Mankuzhiyil}, {Mannheim}, {Maraschi},
  {Marcote}, {Mariotti}, {Mart{\'{\i}}nez}, {Mazin}, {Menzel}, {Miranda},
  {Mirzoyan}, {Moralejo}, {Munar-Adrover}, {Nakajima}, {Niedzwiecki},
  {Nilsson}, {Nishijima}, {Noda}, {Orito}, {Overkemping}, {Paiano},
  {Palatiello}, {Paneque}, {Paoletti}, {Paredes}, {Paredes-Fortuny}, {Persic},
  {Prada Moroni}, {Prandini}, {Puljak}, {Reinthal}, {Rhode}, {Rib{\'o}},
  {Rico}, {Rodriguez Garcia}, {R{\"u}gamer}, {Saito}, {Saito}, {Satalecka},
  {Scalzotto}, {Scapin}, {Schultz}, {Schweizer}, {Sun}, {Shore},
  {Sillanp{\"a}{\"a}}, {Sitarek}, {Snidaric}, {Sobczynska}, {Spanier},
  {Stamatescu}, {Stamerra}, {Steinbring}, {Steinke}, {Storz}, {Strzys},
  {Takalo}, {Takami}, {Tavecchio}, {Temnikov}, {Terzi{\'c}}, {Tescaro},
  {Teshima}, {Thaele}, {Tibolla}, {Torres}, {Toyama}, {Treves}, {Uellenbeck},
  {Vogler}, \& {Zanin}}]{2015A&A...578A..22A}
{Aleksi{\'c}}, J., {Ansoldi}, S., {Antonelli}, L.~A., {et~al.}
  2015{\natexlab{b}}, \aap, 578, A22

\bibitem[{{Aleksi{\'c}} {et~al.}(2015{\natexlab{c}}){Aleksi{\'c}}, {Ansoldi},
  {Antonelli}, {Antoranz}, {Babic}, {Bangale}, {Barres de Almeida}, {Barrio},
  {Becerra Gonz{\'a}lez}, {Bednarek}, \& et~al.}]{2015A&A...573A..50A}
{Aleksi{\'c}}, J., {Ansoldi}, S., {Antonelli}, L.~A., {et~al.}
  2015{\natexlab{c}}, \aap, 573, A50

\bibitem[{{Aliu} {et~al.}(2009){Aliu}, {Anderhub}, {Antonelli}, {Antoranz},
  {Backes}, {Baixeras}, {Barrio}, {Bartko}, {Bastieri}, {Becker}, {Bednarek},
  {Berger}, {Bernardini}, {Biland}, {Bock}, {Bonnoli}, {Bordas}, {Borla
  Tridon}, {Bosch-Ramon}, {Bretz}, {Britvitch}, {Camara}, {Carmona},
  {Chilingarian}, {Commichau}, {Contreras}, {Cortina}, {Costado}, {Covino},
  {Curtef}, {Dazzi}, {de Angelis}, {de Cea Del Pozo}, {de Los Reyes}, {de
  Lotto}, {de Maria}, {de Sabata}, {Delgado Mendez}, {Dominguez}, {Dorner},
  {Doro}, {Els{\"a}sser}, {Errando}, {Fagiolini}, {Ferenc}, {Fern{\'a}ndez},
  {Firpo}, {Fonseca}, {Font}, {Galante}, {Garc{\'{\i}}a L{\'o}pez},
  {Garczarczyk}, {Gaug}, {Goebel}, {Hadasch}, {Hayashida}, {Herrero},
  {H{\"o}hne}, {Hose}, {Hsu}, {Huber}, {Jogler}, {Kranich}, {La Barbera},
  {Laille}, {Leonardo}, {Lindfors}, {Lombardi}, {Longo}, {L{\'o}pez}, {Lorenz},
  {Majumdar}, {Maneva}, {Mankuzhiyil}, {Mannheim}, {Maraschi}, {Mariotti},
  {Mart{\'{\i}}nez}, {Mazin}, {Meucci}, {Meyer}, {Miranda}, {Mirzoyan},
  {Moles}, {Moralejo}, {Nieto}, {Nilsson}, {Ninkovic}, {Otte}, {Oya},
  {Paoletti}, {Paredes}, {Pasanen}, {Pascoli}, {Pauss}, {Pegna},
  {Perez-Torres}, {Persic}, {Peruzzo}, {Piccioli}, {Prada}, {Prandini},
  {Puchades}, {Raymers}, {Rhode}, {Rib{\'o}}, {Rico}, {Rissi}, {Robert},
  {R{\"u}gamer}, {Saggion}, {Saito}, {Salvati}, {Sanchez-Conde}, {Sartori},
  {Satalecka}, {Scalzotto}, {Scapin}, {Schweizer}, {Shayduk}, {Shinozaki},
  {Shore}, {Sidro}, {Sierpowska-Bartosik}, {Sillanp{\"a}{\"a}}, {Sitarek},
  {Sobczynska}, {Spanier}, {Stamerra}, {Stark}, {Takalo}, {Tavecchio},
  {Temnikov}, {Tescaro}, {Teshima}, {Tluczykont}, {Torres}, {Turini}, {Vankov},
  {Venturini}, {Vitale}, {Wagner}, {Wittek}, {Zabalza}, {Zandanel}, {Zanin}, \&
  {Zapatero}}]{2009APh....30..293A}
{Aliu}, E., {Anderhub}, H., {Antonelli}, L.~A., {et~al.} 2009, Astroparticle
  Physics, 30, 293

\bibitem[{{Balokovi{\'c}} {et~al.}(2016){Balokovi{\'c}}, {Paneque}, {Madejski},
  {Furniss}, {Chiang}, {Ajello}, {Alexander}, {Barret}, {Blandford}, {Boggs},
  \& et~al.}]{2016ApJ...819..156B}
{Balokovi{\'c}}, M., {Paneque}, D., {Madejski}, G., {et~al.} 2016, \apj, 819,
  156

\bibitem[{{Bartoli} {et~al.}(2011){Bartoli}, {Bernardini}, {Bi}, {Bleve},
  {Bolognino}, {Branchini}, {Budano}, {Calabrese Melcarne}, {Camarri}, {Cao},
  {Cappa}, {Cardarelli}, {Catalanotti}, {Cattaneo}, {Celio}, {Chen}, {Chen},
  {Chen}, {Creti}, {Cui}, {Dai}, {D'Al{\'{\i}} Staiti}, {Danzengluobu},
  {Dattoli}, {De Mitri}, {D'Ettorre Piazzoli}, {Di Girolamo}, {Ding}, {Di
  Sciascio}, {Feng}, {Feng}, {Feng}, {Galeazzi}, {Galeotti}, {Giroletti},
  {Gou}, {Guo}, {He}, {Hu}, {Hu}, {Huang}, {Iacovacci}, {Iuppa}, {James},
  {Jia}, {Labaciren}, {Li}, {Li}, {Li}, {Liguori}, {Liu}, {Liu}, {Liu}, {Liu},
  {Lu}, {Ma}, {Mancarella}, {Mari}, {Marsella}, {Martello}, {Mastroianni},
  {Montini}, {Ning}, {Pagliaro}, {Panareo}, {Panico}, {Perrone}, {Pistilli},
  {Qu}, {Rossi}, {Ruggieri}, {Salvini}, {Santonico}, {Shen}, {Sheng}, {Shi},
  {Stanescu}, {Surdo}, {Tan}, {Vallania}, {Vernetto}, {Vigorito}, {Wang},
  {Wang}, {Wu}, {Wu}, {Xu}, {Xue}, {Yan}, {Yang}, {Yang}, {Yao}, {Yuan}, {Zha},
  {Zhang}, {Zhang}, {Zhang}, {Zhang}, {Zhang}, {Zhang}, {Zhang}, {Zhaxiciren},
  {Zhaxisangzhu}, {Zhou}, {Zhu}, {Zhu}, {Zizzi}, \& {ARGO-YBJ
  Collaboration}}]{2011ApJ...734..110B}
{Bartoli}, B., {Bernardini}, P., {Bi}, X.~J., {et~al.} 2011, \apj, 734, 110

\bibitem[{{B{\l}a{\.z}ejowski} {et~al.}(2005){B{\l}a{\.z}ejowski}, {Blaylock},
  {Bond}, {Bradbury}, {Buckley}, {Carter-Lewis}, {Celik}, {Cogan}, {Cui},
  {Daniel}, {Duke}, {Falcone}, {Fegan}, {Fegan}, {Finley}, {Fortson},
  {Gammell}, {Gibbs}, {Gillanders}, {Grube}, {Gutierrez}, {Hall}, {Hanna},
  {Holder}, {Horan}, {Humensky}, {Kenny}, {Kertzman}, {Kieda}, {Kildea},
  {Knapp}, {Kosack}, {Krawczynski}, {Krennrich}, {Lang}, {LeBohec}, {Linton},
  {Lloyd-Evans}, {Maier}, {Mendoza}, {Milovanovic}, {Moriarty}, {Nagai}, {Ong},
  {Power-Mooney}, {Quinn}, {Quinn}, {Ragan}, {Reynolds}, {Rebillot}, {Rose},
  {Schroedter}, {Sembroski}, {Swordy}, {Syson}, {Valcarel}, {Vassiliev},
  {Wakely}, {Walker}, {Weekes}, {White}, {Zweerink}, {VERITAS Collaboration},
  {Mochejska}, {Smith}, {Aller}, {Aller}, {Ter{\"a}sranta}, {Boltwood},
  {Sadun}, {Stanek}, {Adams}, {Foster}, {Hartman}, {Lai}, {B{\"o}ttcher},
  {Reimer}, \& {Jung}}]{2005ApJ...630..130B}
{B{\l}a{\.z}ejowski}, M., {Blaylock}, G., {Bond}, I.~H., {et~al.} 2005, \apj,
  630, 130

\bibitem[{{Buckley} {et~al.}(1996){Buckley}, {Akerlof}, {Biller},
  {Carter-Lewis}, {Catanese}, {Cawley}, {Connaughton}, {Fegan}, {Finley},
  {Gaidos}, {Hillas}, {Kartje}, {Koenigl}, {Krennrich}, {Lamb}, {Lessard},
  {Macomb}, {Mattox}, {McEnery}, {Mohanty}, {Quinn}, {Rodgers}, {Rose},
  {Schubnel}, {Sembroski}, {Smith}, {Weekes}, {Wilson}, \&
  {Zweerink}}]{1996ApJ...472L...9B}
{Buckley}, J.~H., {Akerlof}, C.~W., {Biller}, S., {et~al.} 1996, \apjl, 472, L9

\bibitem[{{Cao} \& {Wang}(2013)}]{2013PASJ...65..109C}
{Cao}, G. \& {Wang}, J. 2013, \pasj, 65

\bibitem[{{Chatterjee} {et~al.}(2012){Chatterjee}, {Bailyn}, {Bonning},
  {Buxton}, {Coppi}, {Fossati}, {Isler}, {Maraschi}, \&
  {Urry}}]{2012ApJ...749..191C}
{Chatterjee}, R., {Bailyn}, C.~D., {Bonning}, E.~W., {et~al.} 2012, \apj, 749,
  191

\bibitem[{{Chatterjee} {et~al.}(2008){Chatterjee}, {Jorstad}, {Marscher}, {Oh},
  {McHardy}, {Aller}, {Aller}, {Balonek}, {Miller}, {Ryle}, {Tosti},
  {Kurtanidze}, {Nikolashvili}, {Larionov}, \&
  {Hagen-Thorn}}]{2008ApJ...689...79C}
{Chatterjee}, R., {Jorstad}, S.~G., {Marscher}, A.~P., {et~al.} 2008, \apj,
  689, 79

\bibitem[{{Cui}(2004)}]{2004ApJ...605..662C}
{Cui}, W. 2004, \apj, 605, 662

\bibitem[{{Donnarumma} {et~al.}(2009){Donnarumma}, {Vittorini}, {Vercellone},
  {del Monte}, {Feroci}, {D'Ammando}, {Pacciani}, {Chen}, {Tavani},
  {Bulgarelli}, \& et~al.}]{2009ApJ...691L..13D}
{Donnarumma}, I., {Vittorini}, V., {Vercellone}, S., {et~al.} 2009, \apjl, 691,
  L13

\bibitem[{{Edelson} \& {Krolik}(1988)}]{1988ApJ...333..646E}
{Edelson}, R.~A. \& {Krolik}, J.~H. 1988, \apj, 333, 646

\bibitem[{{Fomin} {et~al.}(1994){Fomin}, {Stepanian}, {Lamb}, {Lewis}, {Punch},
  \& {Weekes}}]{1994APh.....2..137F}
{Fomin}, V.~P., {Stepanian}, A.~A., {Lamb}, R.~C., {et~al.} 1994, Astroparticle
  Physics, 2, 137

\bibitem[{{Fossati} {et~al.}(2004){Fossati}, {Buckley}, {Edelson}, {Horns}, \&
  {Jordan}}]{2004NewAR..48..419F}
{Fossati}, G., {Buckley}, J., {Edelson}, R.~A., {Horns}, D., \& {Jordan}, M.
  2004, \nar, 48, 419

\bibitem[{{Fossati} {et~al.}(2008){Fossati}, {Buckley}, {Bond}, {Bradbury},
  {Carter-Lewis}, {Chow}, {Cui}, {Falcone}, {Finley}, {Gaidos}, {Grube},
  {Holder}, {Horan}, {Horns}, {Jordan}, {Kieda}, {Kildea}, {Krawczynski},
  {Krennrich}, {Lang}, {LeBohec}, {Lee}, {Moriarty}, {Ong}, {Petry}, {Quinn},
  {Sembroski}, {Wakely}, \& {Weekes}}]{2008ApJ...677..906F}
{Fossati}, G., {Buckley}, J.~H., {Bond}, I.~H., {et~al.} 2008, \apj, 677, 906

\bibitem[{{Gaidos} {et~al.}(1996){Gaidos}, {Akerlof}, {Biller}, {Boyle},
  {Breslin}, {Buckley}, {Carter-Lewis}, {Catanese}, {Cawley}, {Fegan},
  {Finley}, {Gordo}, {Hillas}, {Krennrich}, {Lamb}, {Lessard}, {McEnery},
  {Masterson}, {Mohanty}, {Moriarty}, {Quinn}, {Rodgers}, {Rose}, {Samuelson},
  {Schubnell}, {Sembroski}, {Srinivasan}, {Weekes}, {Wilson}, \&
  {Zweerink}}]{1996Natur.383..319G}
{Gaidos}, J.~A., {Akerlof}, C.~W., {Biller}, S., {et~al.} 1996, \nat, 383, 319

\bibitem[{{Giebels} {et~al.}(2007){Giebels}, {Dubus}, \&
  {Kh{\'e}lifi}}]{2007A&A...462...29G}
{Giebels}, B., {Dubus}, G., \& {Kh{\'e}lifi}, B. 2007, \aap, 462, 29

\bibitem[{{Horan} {et~al.}(2009){Horan}, {Acciari}, {Bradbury}, {Buckley},
  {Bugaev}, {Byrum}, {Cannon}, {Celik}, {Cesarini}, {Chow}, {Ciupik}, {Cogan},
  {Falcone}, {Fegan}, {Finley}, {Fortin}, {Fortson}, {Gall}, {Gillanders},
  {Grube}, {Gyuk}, {Hanna}, {Hays}, {Kertzman}, {Kildea}, {Konopelko},
  {Krawczynski}, {Krennrich}, {Lang}, {Lee}, {Moriarty}, {Nagai}, {Niemiec},
  {Ong}, {Perkins}, {Pohl}, {Quinn}, {Reynolds}, {Rose}, {Sembroski}, {Smith},
  {Steele}, {Swordy}, {Toner}, {Vassiliev}, {Wakely}, {Weekes}, {White},
  {Williams}, {Wood}, {Zitzer}, {Aller}, {Aller}, {Baker}, {Barnaby}, {Carini},
  {Charlot}, {Dumm}, {Fields}, {Hovatta}, {Jordan}, {Kovalev}, {Kovalev},
  {Krimm}, {Kurtanidze}, {L{\"a}hteenm{\"a}ki}, {LeCampion}, {Maune},
  {Montaruli}, {Sadun}, {Smith}, {Tornikoski}, {Turunen}, \&
  {Walters}}]{2009ApJ...695..596H}
{Horan}, D., {Acciari}, V.~A., {Bradbury}, S.~M., {et~al.} 2009, \apj, 695, 596

\bibitem[{{Hovatta} {et~al.}(2015){Hovatta}, {Petropoulou}, {Richards},
  {Giannios}, {Wiik}, {Balokovi{\'c}}, {L{\"a}hteenm{\"a}ki}, {Lott},
  {Max-Moerbeck}, {Ramakrishnan}, \& {Readhead}}]{2015MNRAS.448.3121H}
{Hovatta}, T., {Petropoulou}, M., {Richards}, J.~L., {et~al.} 2015, \mnras,
  448, 3121

\bibitem[{{Katarzy{\'n}ski} {et~al.}(2003){Katarzy{\'n}ski}, {Sol}, \&
  {Kus}}]{2003A&A...410..101K}
{Katarzy{\'n}ski}, K., {Sol}, H., \& {Kus}, A. 2003, \aap, 410, 101

\bibitem[{{Krimm} {et~al.}(2013){Krimm}, {Holland}, {Corbet}, {Pearlman},
  {Romano}, {Kennea}, {Bloom}, {Barthelmy}, {Baumgartner}, {Cummings},
  {Gehrels}, {Lien}, {Markwardt}, {Palmer}, {Sakamoto}, {Stamatikos}, \&
  {Ukwatta}}]{2013ApJS..209...14K}
{Krimm}, H.~A., {Holland}, S.~T., {Corbet}, R.~H.~D., {et~al.} 2013, \apjs,
  209, 14

\bibitem[{{Levine} {et~al.}(1996){Levine}, {Bradt}, {Cui}, {Jernigan},
  {Morgan}, {Remillard}, {Shirey}, \& {Smith}}]{1996ApJ...469L..33L}
{Levine}, A.~M., {Bradt}, H., {Cui}, W., {et~al.} 1996, \apjl, 469, L33

\bibitem[{{Lico} {et~al.}(2014){Lico}, {Giroletti}, {Orienti}, {G{\'o}mez},
  {Casadio}, {D'Ammando}, {Blasi}, {Cotton}, {Edwards}, {Fuhrmann}, {Jorstad},
  {Kino}, {Kovalev}, {Krichbaum}, {Marscher}, {Paneque}, {Piner}, \&
  {Sokolovsky}}]{2014A&A...571A..54L}
{Lico}, R., {Giroletti}, M., {Orienti}, M., {et~al.} 2014, \aap, 571, A54

\bibitem[{{Macomb} {et~al.}(1995){Macomb}, {Akerlof}, {Aller}, {Aller},
  {Bertsch}, {Bruhweiler}, {Buckley}, {Carter-Lewis}, {Cawley}, {Cheng},
  {Dermer}, {Fegan}, {Gaidos}, {Gear}, {Hall}, {Hartman}, {Hillas}, {Kafatos},
  {Kerrick}, {Kniffen}, {Kondo}, {Kubo}, {Lamb}, {Makino}, {Makishima},
  {Marscher}, {McEnery}, {McHardy}, {Meyer}, {Moore}, {Ramos}, {Robson},
  {Rose}, {Schubnell}, {Sembroski}, {Stevens}, {Takahashi}, {Tashiro},
  {Weekes}, {Wilson}, \& {Zweerink}}]{1995ApJ...449L..99M}
{Macomb}, D.~J., {Akerlof}, C.~W., {Aller}, H.~D., {et~al.} 1995, \apjl, 449,
  L99

\bibitem[{{Mannheim}(1993)}]{1993A&A...269...67M}
{Mannheim}, K. 1993, \aap, 269, 67

\bibitem[{{Max-Moerbeck} {et~al.}(2014){Max-Moerbeck}, {Hovatta}, {Richards},
  {King}, {Pearson}, {Readhead}, {Reeves}, {Shepherd}, {Stevenson},
  {Angelakis}, {Fuhrmann}, {Grainge}, {Pavlidou}, {Romani}, \&
  {Zensus}}]{2014MNRAS.445..428M}
{Max-Moerbeck}, W., {Hovatta}, T., {Richards}, J.~L., {et~al.} 2014, \mnras,
  445, 428

\bibitem[{{Nilsson} {et~al.}(2007){Nilsson}, {Pasanen}, {Takalo}, {Lindfors},
  {Berdyugin}, {Ciprini}, \& {Pforr}}]{2007A&A...475..199N}
{Nilsson}, K., {Pasanen}, M., {Takalo}, L.~O., {et~al.} 2007, \aap, 475, 199

\bibitem[{{Piner} {et~al.}(1999){Piner}, {Unwin}, {Wehrle}, {Edwards}, {Fey},
  \& {Kingham}}]{1999ApJ...525..176P}
{Piner}, B.~G., {Unwin}, S.~C., {Wehrle}, A.~E., {et~al.} 1999, \apj, 525, 176

\bibitem[{{Poutanen} {et~al.}(2008){Poutanen}, {Zdziarski}, \&
  {Ibragimov}}]{2008MNRAS.389.1427P}
{Poutanen}, J., {Zdziarski}, A.~A., \& {Ibragimov}, A. 2008, \mnras, 389, 1427

\bibitem[{{Punch} {et~al.}(1992){Punch}, {Akerlof}, {Cawley}, {Chantell},
  {Fegan}, {Fennell}, {Gaidos}, {Hagan}, {Hillas}, {Jiang}, {Kerrick}, {Lamb},
  {Lawrence}, {Lewis}, {Meyer}, {Mohanty}, {O'Flaherty}, {Reynolds}, {Rovero},
  {Schubnell}, {Sembroski}, {Weekes}, \& {Wilson}}]{1992Natur.358..477P}
{Punch}, M., {Akerlof}, C.~W., {Cawley}, M.~F., {et~al.} 1992, \nat, 358, 477

\bibitem[{{Richards} {et~al.}(2011){Richards}, {Max-Moerbeck}, {Pavlidou},
  {King}, {Pearson}, {Readhead}, {Reeves}, {Shepherd}, {Stevenson},
  {Weintraub}, {Fuhrmann}, {Angelakis}, {Zensus}, {Healey}, {Romani}, {Shaw},
  {Grainge}, {Birkinshaw}, {Lancaster}, {Worrall}, {Taylor}, {Cotter}, \&
  {Bustos}}]{2011ApJS..194...29R}
{Richards}, J.~L., {Max-Moerbeck}, W., {Pavlidou}, V., {et~al.} 2011, \apjs,
  194, 29

\bibitem[{{Scargle}(1998)}]{1998ApJ...504..405S}
{Scargle}, J.~D. 1998, \apj, 504, 405

\bibitem[{{Scargle} {et~al.}(2013){Scargle}, {Norris}, {Jackson}, \&
  {Chiang}}]{2013ApJ...764..167S}
{Scargle}, J.~D., {Norris}, J.~P., {Jackson}, B., \& {Chiang}, J. 2013, \apj,
  764, 167

\bibitem[{{Schlafly} \& {Finkbeiner}(2011)}]{2011ApJ...737..103S}
{Schlafly}, E.~F. \& {Finkbeiner}, D.~P. 2011, \apj, 737, 103

\bibitem[{{Takalo} {et~al.}(2008){Takalo}, {Nilsson}, {Lindfors},
  {Sillanp{\"a}{\"a}}, {Berdyugin}, \& {Pasanen}}]{2008AIPC.1085..705T}
{Takalo}, L.~O., {Nilsson}, K., {Lindfors}, E., {et~al.} 2008, in American
  Institute of Physics Conference Series, Vol. 1085, American Institute of
  Physics Conference Series, ed. F.~A. {Aharonian}, W.~{Hofmann}, \&
  F.~{Rieger}, 705--707

\bibitem[{{Ter\"asranta} {et~al.}(1998){Ter\"asranta}, {Tornikoski}, {Mujunen},
  {Karlamaa}, {Valtonen}, {Henelius}, {Urpo}, {Lainela}, {Pursimo}, {Nilsson},
  {Wiren}, {Laehteenmaeki}, {Korpi}, {Rekola}, {Heinaemaeki}, {Hanski},
  {Nurmi}, {Kokkonen}, {Keinaenen}, {Joutsamo}, {Oksanen}, {Pietilae},
  {Valtaoja}, {Valtonen}, \& {Koenoenen}}]{1998A&AS..132..305T}
{Ter\"asranta}, H., {Tornikoski}, M., {Mujunen}, A., {et~al.} 1998, \aaps, 132,
  305

\bibitem[{{Tluczykont} {et~al.}(2010){Tluczykont}, {Bernardini}, {Satalecka},
  {Clavero}, {Shayduk}, \& {Kalekin}}]{2010A&A...524A..48T}
{Tluczykont}, M., {Bernardini}, E., {Satalecka}, K., {et~al.} 2010, \aap, 524,
  A48

\bibitem[{{Uttley} {et~al.}(2003){Uttley}, {Edelson}, {McHardy}, {Peterson}, \&
  {Markowitz}}]{2003ApJ...584L..53U}
{Uttley}, P., {Edelson}, R., {McHardy}, I.~M., {Peterson}, B.~M., \&
  {Markowitz}, A. 2003, \apjl, 584, L53

\bibitem[{{Vaughan} {et~al.}(2003){Vaughan}, {Edelson}, {Warwick}, \&
  {Uttley}}]{2003MNRAS.345.1271V}
{Vaughan}, S., {Edelson}, R., {Warwick}, R.~S., \& {Uttley}, P. 2003, \mnras,
  345, 1271

\bibitem[{{Villata} {et~al.}(2008){Villata}, {Raiteri}, {Larionov},
  {Kurtanidze}, {Nilsson}, {Aller}, {Tornikoski}, {Volvach}, {Aller},
  {Arkharov}, {Bach}, {Beltrame}, {Bhatta}, {Buemi}, {B{\"o}ttcher},
  {Calcidese}, {Carosati}, {Castro-Tirado}, {da Rio}, {di Paola}, {Dolci},
  {Forn{\'e}}, {Frasca}, {Hagen-Thorn}, {Heidt}, {Hiriart}, {Jel{\'{\i}}nek},
  {Kimeridze}, {Konstantinova}, {Kopatskaya}, {Lanteri}, {Leto}, {Ligustri},
  {Lindfors}, {L{\"a}hteenm{\"a}ki}, {Marilli}, {Nieppola}, {Nikolashvili},
  {Pasanen}, {Ragozzine}, {Ros}, {Sigua}, {Smart}, {Sorcia}, {Takalo},
  {Tavani}, {Trigilio}, {Turchetti}, {Uckert}, {Umana}, {Vercellone}, \&
  {Webb}}]{2008A&A...481L..79V}
{Villata}, M., {Raiteri}, C.~M., {Larionov}, V.~M., {et~al.} 2008, \aap, 481,
  L79

\bibitem[{{Zanin} {et~al.}(2013){Zanin}, {Carmona}, {Sitarek}, {Colin},
  {Frantzen}, {Gaug}, {Lombardi}, {Lopez}, {Moralejo}, {Satalecka}, {Scapin},
  {Stamatescu}, \& {for the MAGIC collaboration}}]{Zanin2013}
{Zanin}, R., {Carmona}, E., {Sitarek}, J., {et~al.} 2013, in Proceedings of
  33rd International Cosmic Ray Conference (Rio de Janeiro, Brazil)

\end{thebibliography}

\end{document}